\documentclass[aps,prb,twocolumn,showpacs,showkeys]{revtex4-1}
\usepackage{bm}
\usepackage{ulem}
\usepackage{ctable}
\bibpunct{}{}{,}{s}{}{}
\usepackage{ulem}
\begin{document}
\title{Electronic structure of Pr$_{1-x}$Ca$_{x}$MnO$_3$}
\author{Mohsen Sotoudeh}
\author{Sangeeta Rajpurohit}
\author{Peter Bl\"ochl}
\email[corresponding author: ]{peter.bloechl@tu-clausthal.de}
\affiliation{Institute for Theoretical Physics, Clausthal University
  of Technology, Leibnizstr. 10, 38678 Clausthal-Zellerfeld, Germany }
\author{Daniel Mierwaldt}
\author{Jonas Norpoth}
\author{Vladimir Roddatis} 
\author{Stephanie Mildner} 
\author{Birte Kressdorf}
\author{Benedikt Ifland} 
\author{Christian Jooss}
\affiliation{Institut f\"ur Materialphysik, 
Universit\"at G\"ottingen, Friedrich-Hund-Platz 1,
  37077 G\"ottingen, Germany}
\date {\today}

\begin{abstract}
The electronic structure of {Pr$_{1-x}$Ca$_{x}$MnO$_3$} has been
investigated using a combination of first-principles calculations,
X-ray photoelectron spectroscopy (XPS), X-ray absorption spectroscopy
(XAS), electron-energy loss spectroscopy (EELS), and optical
absorption. The full range of compositions, $x=0,1/2,1$, and a variety
of magnetic orders have been covered. Jahn-Teller as well as Zener
polaron orders are considered. The free parameters of the local hybrid
density functionals used in this study has been determined by
comparison with measured XPS spectra.  A model Hamiltonian, valid for
the entire doping range, has been extracted.  A simple local-orbital
picture of the electronic structure for the interpretation of
experimental spectra is provided. The comparison of theoretical
calculations and different experimental spectra provide a detailed
and consistent picture of the electronic structure.  The large
variations of measured optical absorption spectra are traced back to
the coexistence of magnetic orders respectively to the occupation of
local orbitals.  A consistent treatment of the Coulomb interaction
indicate a partial cancellation of Coulomb parameters and support the
dominance of the electron-phonon coupling.
\end{abstract}

\maketitle

\section{Introduction}
Manganites have attracted wide interest due to their strong correlations
between charge, spin, orbital and lattice degrees of freedom.  These
correlations lead to a rich phase diagram including two-phase regions
with nano-scale phase separation.~\cite{tokura06_repprogphys69_797,
  imada98_rmp70_1039, jirak85_jmmm53_153,jooss07_pnas104_13597,
  tomioka96_prb53_R1689} In addition, manganites are well known for
their colossal magnetoresistance (CMR)
effect\cite{salamon01_rmp73_583,dagotto01_physrep344_1,dagotto02_book}
and the emergence of exotic phases including
charge-ordered,\cite{goodenough55_pr100_564}
multiferroic\cite{jooss07_pnas104_13597} and pseudogap
states.\cite{manella05_nature438_474} They may be used to design more
efficient correlated-electron
devices,~\cite{rozenberg06_apl88_033510,molinari14_jmca2_14109} e.g.
in magnetoelectronics and electronic memory elements with smaller
size.

Recently, interest in doped manganites arose from their exotic
photovoltaic properties\cite{dho10_ssc150_2243,saucke12_prb85_165315}
as well as their potential benefit as
electrocatalysts.\cite{raabe12_advfunctmater22_3378} The latter is
inspired by the use of a CaMn$_4$O$_5$ complex as oxygen-evolution
center in natural photosynthesis.\cite{ferreira04_science303_1831}
Theoretical work on descriptors for catalytic oxygen-evolution
activity rely on a detailed knowledge of the electronic properties of
Mn-O bonds in manganite
perovskites.\cite{rossmeisl07_jeac607_83,man11_chemcatchem3_1159}
While arguments based on electron filling of a rigid antibonding state
between Mn-3d and O-2p orbitals describe some trends in oxygen
evolution activity,\cite{suntivich11_science334_1383} they do not
cover all observed trends.\cite{hong15_energyenvironmentalscience8_1404}

Doping, which determines the charge state of the Mn ion, is the main
control parameter within the complex phase diagram of manganites.  A
wide range of other parameters such as ion
sizes,\cite{ramirez97_jpcm9_8171,imada98_rmp70_1039} dimensionality,
etc. allows to tune the phase boundaries.  In addition, external
fields and temperature allow to actively drive the system across a
phase boundary, which radically changes its properties and thus holds
promise for technological applicability.

The series of {Pr$_{1-x}$Ca$_{x}$MnO$_3$} is of special interest
because of the similar ion sizes of Pr and Ca. They allow to change
the charge of the Mn ion over the entire range from Mn$^{3+}$ to
Mn$^{4+}$ without greatly affecting the crystal
structure.~\cite{stankiewicz00_prb61_11236,jooss07_pnas104_13597}

Early electronic structure calculations\cite{satpathy96_prl76_960} for
manganites indicated that conventional density functionals do not
capture the electronic structure in their electron spectrum. This
finding is in line with that for other transition-metal oxides.  It
has been shown, that the spectral properties can be improved by the
so-called LDA+U method,\cite{anisimov91_prb44_943} which adds an
explicit Fock term and thus introduces the atomic physics of strongly
interacting electrons.  A more rigorous basis for this description has
been given by the hybrid density
functionals.\cite{becke93_jcp98_1372,heyd03_jcp118_8207} The
one-particle spectrum extracted from density-functional calculations,
the so-called Kohn-Sham spectrum, should, in principle, not be
identified with an excitation spectrum. However, the hybrid
functionals, despite their roots in density-functional theory, have a
close similarity with many-particle Green's function methods such as
the GW method.\cite{hedin65_pr139_796,franchini12_jpcm24_235602} This
analogy provides a conceptional connection of the resulting Kohn-Sham
spectra obtained in density-functional theory with the optical
excitation spectra. In this paper, we exploit this analogy to
  determine the free parameters of the hybrid functionals by
  comparison with experimental spectra.

Most ab-initio calculations focussed on the electron-rich Mn$^{3+}$
systems such as LaMnO$_3$, which is an orbital ordered Jahn-Teller
insulator with a band gap slightly above
1~eV.\cite{kovacik16_prb93_75139, franchini14_jpcm26_253202} Ederer,
Kovacic and collaborators\cite{ederer07_prb76_155105,
  kovacik10_prb81_245108,
  kovacik11_prb84_75118,franchini12_jpcm24_235602,he12_prb85_195135,
  he12_prb86_235117, kovacik16_prb93_75139, franchini14_jpcm26_253202}
performed a careful analysis using hybrid functionals on the
iso-electronic series by exchanging the A-type cation and extracted
parameters for model hamiltonians.

However, a comprehensive comparison of the occupied and empty states
as well as the optical excitation spectrum with experimental data is
still missing.

This is even more the case for a full doping series, where the effect
of doping-induced changes of the correlations on the electronic
structure, e.g. due to charge ordering, must be captured by ab-initio
studies in a fully consistent approach.

The nature of the charge-ordered state at half doping was under debate
for years. Experiments clearly confirmed the absence of strong charge
disproportionation between different Mn
sites,\cite{grenier04_prb69_134419} the presence of multiferroic
states with electric polarization\cite{jooss07_pnas104_13597} and the
formation of Mn-O-Mn
dimers.\cite{daoudaladine02_prl89_97205,wu07_prb76_174210} The
occurrence of an underlying Zener polaron type of charge ordering was
supported by different theoretical
work.\cite{efremov04_naturematerials3_853, colizzi10_prb82_140101,
  giovannetti09_prl103_37601} Colizzi and
Giovanetti\cite{colizzi10_prb82_140101,giovannetti09_prl103_37601}
investigated the half-doped materials {Pr$_{1/2}$Ca$_{1/2}$MnO$_3$} and
La$_{1/2}$Ca$_{1/2}$MnO$_3$ using the measured atomic structure and
confirmed the observed electric polarization. Efremov et
al.\cite{efremov04_naturematerials3_853} used an extended
double-exchange Hamiltonian for the study of the transition of the
CE-type CO structure to the ordering of Zener polarons.  Finding an
exact relation between the bandstructure close to the Fermi level,
excitation spectrum and the type of order is a very important step
towards the future understanding of quasiparticle excitations in the
different phases, using model Hamiltonians in combination with
time-resolved optical spectroscopy.

In this paper, we examine the series {Pr$_{1-x}$Ca$_{x}$MnO$_3$} for
$x=0,\frac{1}{2},1$ to study the effect of doping on the detailed
structure of the valence and conduction band, magnetic and charge
ordering as well as on the resulting excitation spectra.  We use local
hybrid density-functional calculations, which contain one free
parameter per atom type. This free parameter is determined by
experimental X-ray photoelectron spectroscopy (XPS). The resulting
description is verified by comparison with electron energy-loss
near-edge structure (ELNES) and the X-ray absorption near-edge
structure (XANES)\cite{mierwaldt14_catalysts4_129}, as well as
measurements of optical absorption\cite{mildner15_prb92_35145}. The
density-functional calculations in turn serve to determine a minimal
model, which puts the focus on the mechanisms for the polaron order
and the effects of varying correlations on the states forming the band
gap in manganites.  Finally, the insight on the electronic structure
is used to explain the experimental optical conductivity data.

\section{Methods}
\subsection{First-principles calculations}
The calculations were performed in the framework of density functional
theory (DFT)\cite{hohenberg64_pr136_B864,kohn65_pr140_1133} and the 
projector augmented wave method (PAW).\cite{bloechl94_prb50_17953}

The augmentation of the PAW method includes the $5s,6s,5p,6p,5d,4f$
orbital shells on Pr, the $3s,4s,3p,4p,3d$ orbitals on Ca, the
$4s,4p,3d$ orbitals on Mn and the $2s,2p,3d$ orbitals on O. The
partial waves have been constructed using the nodeless partial
waves,\cite{bloechl12_arxiv1210_5937} which implies that the orbital
shells refer to the sequence of nodeless partial waves rather than
atomic orbitals. The matching radii in units of the covalent radii for
the partial wave construction are $r_c/r_{cov}=0.84,0.77,0.8,0.7$ for
$s,p,d,f$ electrons respectively for Pr, $r_c/r_{cov}=0.6$ for all
angular momenta in Ca, $r_c/r_{cov}=1$ for Mn, and $r_c/r_{cov}=0.85$
for O.

The auxiliary wave function are expanded up to a plane-wave cutoff of
40~Ry and the auxiliary density to a plane-wave cutoff of 80~Ry.

The k-point integration uses a grid with an upper wave-length cutoff
of $40~a_B$ of the Fourier expansion in the reciprocal unit cell. This
translates into a grid with $(4\times4\times3)$ division of the
reciprocal lattice vectors for the $\sqrt{2}\times\sqrt{2}\times 2$
unit cell of the \textit{Pbnm} space group used for {PrMnO$_3$} and
{CaMnO$_3$}. For the calculations of half-doped
{Pr$_{1/2}$Ca$_{1/2}$MnO$_3$}, a larger
$2\sqrt{2}\times2\sqrt{2}\times 2$ unit cell has been used, with a
correspondingly smaller $(2\times2\times3)$ k-point grid.

We used the experimental lattice constants from Poeppelmeier et
al.\cite{poeppelmeier82_jssc45_71} for {CaMnO$_3$}, from Jirak et
al.\cite{jirak85_jmmm53_153} for {Pr$_{1/2}$Ca$_{1/2}$MnO$_3$} and
from Alonso et al.\cite{alonso00_inorgchem39_917} for {PrMnO$_3$}.
Two further calculations have been performed for the half-doped
material {Pr$_{1/2}$Ca$_{1/2}$MnO$_3$}, for which we used the
experimental lattice constants\cite{daoudaladine02_prl89_97205} for
$x=0.4$. The notation regarding $a,b,c$-directions used in the
following refers to the \textit{Pbnm} space group, respectively
\textit{P21nm} space group.

For {Pr$_{1/2}$Ca$_{1/2}$MnO$_3$} we used an alternating (rock-salt)
ordering of A-type ions Ca and Pr. For the CE-type order we performed
also one calculation with layers of Pr and Ca ions alternating in
$c$-direction.

The atomic positions in each unit cell have been relaxed without any
restriction. The electronic structure has been limited to collinear
spin arrangements.

The electronic structure in manganites depend strongly on the type of
magnetic order. Wollan\cite{wollan55_pr100_545} introduced a notation
according to which a perovskite with antiferromagnetically arranged Mn
ions is called G-type. The wave vector of the spin wave is oriented
along the $(111)$ direction.  The ferromagnetically ordered material,
wave vector $(000)$ is called B-type.  The perovskite with
antiferromagnetically coupled chains of ferromagnetically aligned
Mn ions with wave vector $(110)$ is called C-type. The antiferromagnet
with ferromagnetic ab-planes with vector $(001)$ is called A-type.  

Calculations have been performed for the G, A, C, and B-type magnetic
orders for all three compositions with $x=0,\frac{1}{2}$ and $1$.  For
the half-doped material {Pr$_{1/2}$Ca$_{1/2}$MnO$_3$} we also
investigated the CE-type and the E-type magnetic order shown in
figure~\ref{fig:afdimers}. Unless mentioned otherwise, results for
{CaMnO$_3$} ($x=1$) are provided for the G-type order, results for
{Pr$_{1/2}$Ca$_{1/2}$MnO$_3$} ($x=1/2)$) for the CE-type order and for
{PrMnO$_3$} ($x=0$) A-type order.

We used our local hybrid functional PBE0r as density functional. PBE0r
is derived from the PBE0~\cite{adamo99_jcp110_6158} functional which
replaces a fraction of the exchange energy $E_x^{PBE}$ of the
PBE~\cite{perdew96_prl77_3865} functional by the explicit nonlocal
Hartree-Fock exchange energy $E_x^{HF}$.
\begin{equation}
E^{PBE0} = E^{PBE} + a_x (E_x^{HF} - E_x^{PBE})
\end{equation}
In the local version, PBE0r, the Kohn-Sham wave functions are mapped
onto localized tight-binding orbitals and only the onsite exchange
terms of the exchange correction are included. 

The restriction of the exchange correction to local terms is a way of
range separation, which mimics the screening of the long-range part
of the interaction in the spirit of the GW approximation. The
limitation of onsite terms is appropriate for materials such as
transition metal oxides with a strongly localized d-orbitals, while it
is not suitable for the description of materials with strong covalent
bonds.

In contrast to other implementations of local hybrid
functionals,~\cite{novak06_pssb243_563,tran06_prb74_155108} and the
closely related LDA+U method, we include the Fock term on all atoms
and we take into account all tight-binding orbitals that contribute
the valence electrons or the core wave functions. Also the exchange
interaction between core and valence electrons is taken into
account. The rationale for the inclusion of all orbitals is to avoid
unphysical energy shifts between orbitals that are included in the
correction relative to those that are not.

Another important difference to the LDA+U method is the
double-counting correction, which subtracts the local terms of the DFT
exchange $E_x^{PBE}$ corresponding to the terms of the Fock term
($E_x^{HF}$) added. Our choice\cite{bloechl11_prb84_205101} is in the
spirit of density-matrix functional theory.

The commonly used value $a_x$ for the admixture in the PBE0 functional
is $a_x=1/4$.\cite{perdew96_jcp105_9982} In the present paper, we
adjust these parameters such that the experimentally observed spectral
features are well described. While the use of spectral features for
this purpose is illegitimate in the context of density functional
theory, a justification is provided by the connection with
many-particle Green's functions and density-matrix functional
theory\cite{bloechl13_prb88_25139} on the one hand and to the GW
approximation on the other.

The local approximation of the hybrid functionals allows us to choose
the mixing parameter individually for each atom, reflecting an
effective local dielectric constant. The determination of the
parameters is described in detail below. We used a mixing factor
$a_x=0.15$ for Pr, $a_x=0.07$ for Mn and $a_x=0.1$ for O and Ca.

Like the augmentation, also the construction of local orbitals uses
the framework of nodeless partial waves described
elsewhere.\cite{bloechl12_arxiv1210_5937} The local orbitals are
constructed in the spirit of the tight-binding linear muffin-tin
orbital (LMTO) method\cite{andersen84_prl53_2571} by enforcing that
only the nodeless scattering partial waves are admitted on the
neighboring sites.\cite{bloechl12_arxiv1210_5937} These orbitals are
approximated by a one-center expansion and truncated beyond a radius
of two times the covalent radius of that atom.

The orbital set includes states up to $5s,5p,5d,4f$ for Pr,
up to $3s,3p$ for Ca, up to $4s,3p,3d$ for Mn and up to $2s,2p$
for O atoms.

\subsection{Calculation of ELNES and XANES spectra}
ELNES describes that part of the electron-energy loss spectrum (EELS)
which samples the contribution of a core-electron excitation of a
given element to the empty orbitals. XANES is similarly based on the
absorption of X-rays by excitations of core electrons into empty
states.

ELNES spectra describe the energy loss function
\begin{equation}
F(\omega) = -\text{Im}(\epsilon_r^{-1}(\omega))  
\end{equation}
due to the excitation of an core electron. $\epsilon_r(\omega)$ is the
contribution of the corresponding processes to the relative dielectric
constant and $\hbar\omega$ is the energy transferred from the fast
probe electron. In this paper, we study the $K$-edge absorption of
oxygen, where the 1s core level is excited.

The relative dielectric constant is obtained as\cite{wooten72_book}
\begin{eqnarray}
   \epsilon_r(\omega) &=& 1 + \frac{4\pi e^2 N}{4\pi \epsilon_0 m_e} 
\nonumber\\
&& \times \sum_n 
\frac{Q_{n,0}}{(E_n-E_0)^2-(\hbar \omega)^2-i\Gamma
\hbar \omega }
\end{eqnarray}
$E_{0}$ is the energy of the ground state and $E_{n}$ is the energy of
the excited many-particle state with a core hole and the excited
electron.  In practice, the excitation energy $E_{n}-E_{0}\approx
(\epsilon_n-\epsilon_c)$ is determined as the difference between the
energy $\epsilon_n$ of an unoccupied Kohn-Sham state $|\psi_n\rangle$
and the level $\epsilon_c$ of the core state $|\psi_c\rangle$.  The
oscillator strengths $Q_{n,0}$ are obtained as
\begin{eqnarray}
Q_{n,0} = \frac{2m}{\hbar^2}
\sum_c(\epsilon_n-\epsilon_c) \frac{1}{3}
\sum_j\left|\langle\psi_c|x_j|\psi_n\rangle\right|^2\;.
\end{eqnarray}
The sum over states implicitly contains the spin multiplicity,
i.e. every spin state is counted once.  The sum over $c$ is the sum of
core levels from the same shell. The index $c$ specifies the orbital
momentum $m$ and the spin quantum numbers of the core shell. Only
unoccupied band states contribute to the spectrum. $N$ is the number
of oxygen atoms per unit volume. $\Gamma$ is the life-time broadening
of the excitation.

In order to evaluate the matrix elements
$\langle\psi_c|x_j|\psi_n\rangle$ between the core state
$|\psi_c\rangle$ and the band state $|\psi_n\rangle$, we use the
approximate representation of the wave function in terms of
all-electron partial waves $|\phi_n\rangle$, i.e.
\begin{equation}
|\psi_{n}\rangle \approx 
\sum_n|\phi_\alpha\rangle
\langle\tilde{p}_\alpha|\tilde{\psi}_n\rangle
\end{equation}
When we express the dipole operator $x_{j}$ by cubic spherical
harmonics, we can express the matrix elements using
the Gaunt coefficients $C_{L,L^{\prime},L^{\prime{\prime}}}$=
$\int d\Omega\; Y_L^\ast{Y_{L^\prime}Y_{L^{\prime{\prime}}}}$ as

\begin{eqnarray}
|\langle\psi_c|x_j|\psi_n\rangle|^2
&=&\biggl|
\sum\limits_{\alpha} \sqrt{\frac{4\pi}{3}}
C_{L_{c},p_{j},L_{\alpha}}
\nonumber\\
&&\times \int dr\; r^3R_c(r)R_\alpha(r)\;
\langle \tilde{p}_\alpha|\tilde{\psi}_n\rangle
\biggr|^2
\end{eqnarray}
where $R_{\alpha}(r)$ is the radial part of the partial wave 
$\phi_{\alpha}(r)$ = $R_{\alpha}(|r|)Y_{L_\alpha}(r)$,
 and $R_c(r)$ is the analogously defined radial
part of the core level. The index $p_j$ species the
cubic spherical harmonic with $\ell=1$, that points in the
direction of $x_j$.

The life-time effects have been estimated as follows: The life-time
broadening of the electron in the conduction bands has been estimated
by Egerton\cite{egerton07_ultramicroscopy107_565} as
\begin{equation}
\Gamma(\epsilon)=\frac{\hbar\sqrt{2\epsilon/m_e}}{\lambda}
\end{equation}
by the free-electron velocity $\sqrt{2\epsilon/m_e}$ and the inelastic
mean free path $\lambda$. The energy $\epsilon$ is the energy of the
band state relative to the absorption edge. For the inelastic mean
free path we use the interpolating formula of
Seah\cite{seah79_surfinterfaceanal1_2}
\begin{equation}
\lambda=\frac{A}{\epsilon^2}+B\sqrt{\epsilon}
\end{equation}
with the values $A = 641$~eV$^2$nm and
$B=0.096$~nm/$\sqrt{\mathrm{eV}}$ for inorganic compounds from Seah's
analysis. 

The smaller effect is the lifetime of the core hole. We follow
Egerton\cite{egerton07_ultramicroscopy107_565} and obtain the
core-hole lifetime broadening for the K-shell from
\begin{equation}
\Gamma_K=\biggl[-0.285
+0.0216\left(\frac{E_{thr}}{\mathrm{eV}}\right)^{0.472}
\biggr]~\mathrm{eV}
\end{equation}
where $E_{thr}$ is the excitation energy at the absorption threshold.
We use the calculated core-level energy $E_{thr}=-515.3~eV$, so that
$\Gamma_K=0.70$~eV.

In this paper, we use the so-called independent electron approximation
(IPA), which ignores the presence of the core-hole. The only effect of
the core hole considered is a rigid energy shift of the spectrum to
adjust the absorption threshold to experiment.  A more accurate
calculation would use the same analysis, but for a supercell
calculation with one-half of a core hole on the probed atom. The
justification for the factor one-half is Slater's transition state
rule. An even more advanced analysis requires the solution of the
Bethe-Salpether equation.\cite{laskowski10_prb82_205104}

The justification of the IPA in the present work is the limitation on
the O $K$ edge (1s-2p transition) because it is particularly suited for
an interpretation based on partial Density of States (DoS)
calculations given the generally strong screening of the 1s core
hole.\cite{davoli86_prb33_2979} In contrast, the spectral shape of the
Mn $L_{2,3}$ edge would be strongly affected by multiplet effects
arising from the strong wavefunction overlap between the 2p hole and
the 3d electrons.

\subsection{Measurement of ELNES spectra}
EELS were recorded on an FEI Titan 80-300 microscope, equipped with a
GIF Quantum 965 ER spectrometer, operated at 300~kV in scanning
transmission electron microscopy (STEM) mode. An electron
monochromator provided an energy resolution of 150~meV. The
convergence semi-angle $\alpha$ was 10~mrad, the inner acceptance
semi-angle $\beta$ was 22~mrad. Spectra were acquired from epitaxial
thin films of {PrMnO$_3$}, {Pr$_{1/2}$Ca$_{1/2}$MnO$_3$} and
Pr$_{0.05}$Ca$_{0.95}$MnO$_3$ prepared by ion beam sputtering. Details
of the growth parameters and film properties can be found in
ref.~\cite{mildner15_prb92_35145}.  Due to large oxygen-vacancy
concentration and the tendency for phase decomposition in undoped
{CaMnO$_3$} films, we have chosen a slightly doped variant for
representation of the x=1 system and discuss its impact, where
required. TEM specimens were prepared by mechanical polishing followed
by Ar$^+$ ion milling at glancing angles of 6 degrees. A power law
type background has been subtracted and the spectra have been
normalized. Savitzky-Golay smoothening has been applied by fitting
sub-sets of 20 adjacent data points with a 3rd order
polynominal.\cite{savitzki64_analyticalchemistry64_1627}

\subsection{Measurement of XANES and XPS spectra}
XPS and XANES measurements were performed at the ISISS beam line of
the synchrotron facility BESSY II.\cite{Knop-Gericke2009213} Spectra
were taken at room temperature from epitaxial thin films of
{PrMnO$_3$}, {Pr$_{1/2}$Ca$_{1/2}$MnO$_3$}, and
Pr$_{0.2}$Ca$_{0.8}$MnO$_3$ prepared by ion beam sputtering. The
experimental details of the XANES spectra have been published
elsewhere.\cite{mierwaldt14_catalysts4_129} The resonant photoemission
spectra were recorded at 930~eV for the Pr-M resonance and at 642~eV
for the Mn L-resonance. Pre-resonance spectra at 920~eV and 635~eV
respectively as well as a Shirley-type background were
subtracted.\cite{shirley72_prb5_4709} The recorded spectra have been
aligned on the energy axis such that the signal exceeds the baseline
by a factor of two. This places the difference spectra onto a common
energy scale.

\subsection{Measurement of optical absorption spectra}
Temperature-dependent (from $T= 80$ to $T=300 K$) optical-absorption
measurements in a Cary Varian 5e spectrometer with unpolarized light
in a wavelength range of $\lambda=250-3300$~nm as well as
reflectivity-corrected optical-absorption measurements in a Cary
Varian 50 at room temperature for $\lambda=190-1100$~nm were carried
out at Pr$_{1-x}$Ca$_x$MnO$_3$ ($x=0, 0.5, 0.95$) thin films on
transparent MgO substrate with a film thickness of $d=100$~nm. The
transmission $\mathcal{T}$ (vertical beam) and reflectivity
$\mathcal{R}$ measurements were performed in a Dual Beam mode with
$I_{\mathcal{T}/\mathcal{R}}/I_0$. The absorption coefficient $\alpha
(\lambda)$ has been obtained via $\alpha (\lambda)=-\frac{1}{d}\cdot
\ln\left(\mathcal{T}/(1-\mathcal{R})\right)$. For measurements at low
temperatures or at wavelengths above $\lambda>1100$~nm a reflectivity
correction is infeasible, thus $\mathcal{R}=0$ is used for the
determination of $\alpha (\lambda)$.

\subsection{Calculation of optical absorption spectra}
In order to interpret the measured optical absorption spectra we used
the expression Eq.~\ref{eq:absonsite} developed in
appendix~\ref{app:optabs}. It decomposes the optical absorption into
the contributions of transition between pairs of local orbitals. Each
contribution is given by a joint DoS of specific pairs
of orbitals and a trivial frequency-dependent factor.  Thus we can
attribute the features in the measured absorption spectrum to
particular transitions which can be related to the projected DoS. The
orbitals selected will be described in the text.

\section{Results and discussion} 

\subsection{Electronic structure of manganites}
Before discussing the details of our results we review the gross
features of the atomic and the electronic structure of the
{Pr$_{1-x}$Ca$_{x}$MnO$_3$}.

In the perovskite structure, Mn and O ions form a network of
corner-sharing MnO$_6$ octahedra.  In the manganites studied
  here, the size of the A-type ions is sufficiently small so that the
octahedra tilt in an alternating manner to increase the ionic
attraction.

\begin{figure}[!htb]
\includegraphics[width=\linewidth]{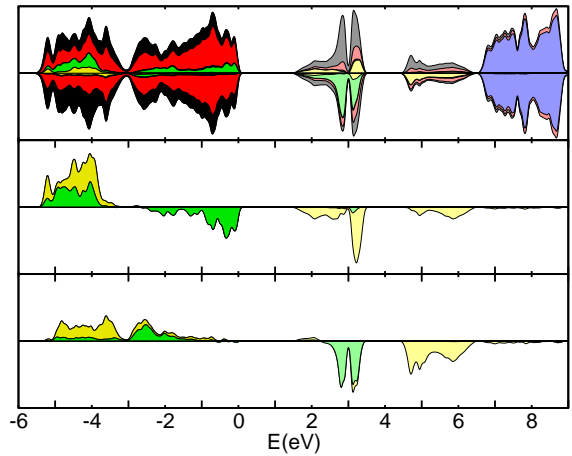}
\caption{\label{fig:cmo_dos}DoS of {CaMnO$_3$} in the stable G-type
  magnetic order. The top figure shows the total DoS (black
  envelope) with the projected DoS for O-p (red), Mn-$t_{2g}$ (green),
  Mn-$e_g$ (yellow), Ca-d (blue). Projected DoS are stacked ontop of
  each other.  The two spin densities are shown with opposite
  sign. The projected DoS is considered only for the Mn ions with one
  majority-spin direction.  The graph in the middle shows the COOP
  between a Mn-$t_{2g}$ orbital and an $\pi$-bonded O-p state in green
  and the COOP between a Mn-$e_g$ orbital and an $\sigma$-bonded
  p-state in yellow for the majority-spin direction. Unlike the DoS,
  the COOPs have positive and negative values, so that two spin
  direction can not be combined into one graph. The two COOPs for
  $t_{2g}$ (green) and $e_g$ (yellow) states are stacked ontop of each
  other. The bottom graph shows the same information for the
  minority-spin direction. The energy zero is aligned with the
  valence-band top. Empty states are drawn with a lighter color than
  filled states.}
\end{figure}

The most simple member of the class of {Pr$_{1-x}$Ca$_{x}$MnO$_3$}
manganites is {CaMnO$_3$}.  The ions are in the formal oxidation states
Ca$^{2+}$Mn$^{4+}$O$^{2-}_3$.

The calculated DoS of {CaMnO$_3$} is shown in figure~\ref{fig:cmo_dos}.
The filled valence band, which extends from -5~eV to 0~eV, is
predominantly of O-p character with some contribution of Mn-d
orbitals. The Ca-d states are located 7-9~eV above the valence band.
The Mn-d states relevant for the complex properties of manganites lie
mostly in between these two features.

The Mn-ions have a large magnetic moment, which leads to a Hund's-rule
splitting between Mn-d levels in the majority- and the minority-spin
direction.

Relevant for the correlations dominating the manganite physics
  are the antibonding states between Mn-d and O-p orbitals. The
  corresponding bonding states are located at the bottom part of the
  oxygen valence band.  Crystal field splitting divides the Mn-d
  states into $t_{2g}$ states and $e_g$ states. We emphasize that the
  crystal-field splitting is not of electrostatic origin but due to
  the covalent interaction with the oxygen
  neighbors.\cite{hotta06_rpp69_2061} Because the $\sigma$ bond of the
  $e_g$ states is stronger than the $\pi$ bonds formed by the $t_{2g}$
  states, the antibonding $e_g$ states lie energetically above the 
  $t_{2g}$ states.

The Crystal-Orbital Overlap Populations (COOPs) in
  Fig.~\ref{fig:cmo_dos} reveal that the three antibonding $t_{2g}$
  states of the majority-spin direction are located in the upper part
  of the oxygen valence band. They are often modelled by a classical
  spin with $S=\frac{3}{2}\hbar$. The majority-spin $e_g$ orbitals in
  CaMnO$_3$ form the conduction band. This band can be filled by
  additional electrons introduced for example by doping. The
  antibonding states of the minority spin are located at higher
  energies, while their DoS overlaps with the majority-spin $e_g$ bands.

Passing from {CaMnO$_3$} to {PrMnO$_3$}, the di-valent Ca ions are replaced
by tri-valent Pr ions and additional electrons are inserted into
anti-bonding $e_g$ states of the Mn sites. These electrons form
polarons. The interplay of magnetic and polaron order are at the
origin of the complex phase diagram of manganites.
\begin{figure}[!htb]
\includegraphics[width=\linewidth]{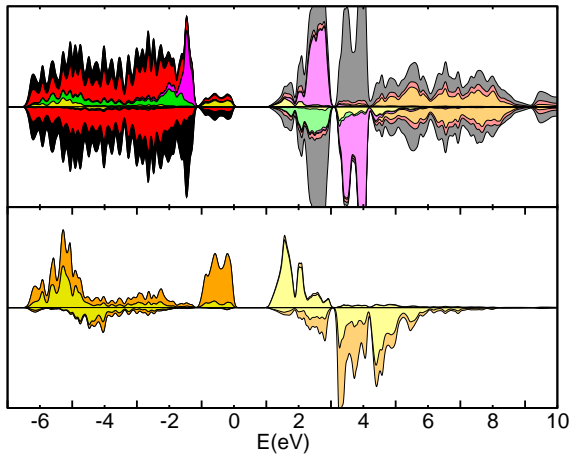}
\caption{\label{fig:pmo_dos}Spin-resolved DoS of {PrMnO$_3$} in the
  stable A-type magnetic order. The color coding in the upper graph is
  described in the text and in figure~\ref{fig:cmo_dos}. The Pr-f
  states are drawn in magenta and the Pr-d states in orange. The graph
  at the bottom shows the DoS for the lower (orange) and upper
  (yellow) $e_g$ orbitals, which are split by the Jahn-Teller effect.}
\end{figure}

The DoS of {PrMnO$_3$} is shown in figure~\ref{fig:pmo_dos}.  The filled
Pr-f states, two per Pr, are located in the upper part of the oxygen
valence band in the range -2~eV to -1~eV.  The empty Pr-f states
overlap energetically with the empty Mn-d states in the range from
2~eV to 4~eV.

In contrast to {CaMnO$_3$}, the band gap in {PrMnO$_3$} is due to the
Jahn-Teller effect: The degeneracy of the Mn-$e_g$ orbitals is lifted
by a distortion of the MnO$_6$ octahedra, which lowers the energy of
the occupied state.  This stabilization of the occupied state is due
to weakening of the antibonds with the neighboring oxygen atoms as a
result of an increase of the corresponding Mn-O bond distances.  The
ordering of the Jahn-Teller distortions give rise to an orbital
ordered state.

All three compounds investigated, with $x=0$, $x=\frac{1}{2}$ and
  $x=1$, are insulating at low temperature. This conflicts with the
  simplified view of carrier doping
  known from semiconductor physics. In the manganites, electrons are
  strongly correlated with spin and lattice degrees of freedom. As a
  consequence, additional electrons tend to form new bound states with
  existing quasi-particles. This reflects in a strong structural and
  magnetic relaxation of the material upon addition or removal of
  electrons. This relaxation is often able to split off the state with
  the additional electron from the conduction band, and to shift it
  into the valence band.

\subsection{Role of the U-tensor}
The electronic spectra of transition metal oxides obtained from
conventional gradient corrected density-functionals such as
PBE\cite{perdew96_prl77_3865} differ appreciably from their excitation
spectra. For example, the band gap of {CaMnO$_3$} in the PBE density
functional is due to a transition from the $t_{2g}$ to $e_g$ states,
while in reality it is a charge-transfer insulator with a transition
from the O-p states to the $e_g$ states.

While an agreement between Kohn-Sham spectra and excitation spectra is
not a theoretical requirement, there are density functionals that
produce spectra which are closer to excitation spectra. Hybrid
functionals incorporate a fraction of an explicit exchange term very
similar to the screened exchange of the GW
method.\cite{hedin65_pr139_796} The GW method, while approximate, has
a sound theoretical basis for predicting the spectrum of one-particle
excitations.

Qualitative differences between Kohn-Sham and excitation spectra are
often a sign for a fundamental flaw of the density functional used,
which also affect the description of the total energy.  This is the
reason for seeking a functional that is able to describe spectral
properties in addition to giving a correct description of the
energetics.

The main effect of admixing the non-local expression for the exchange
in the hybrid functionals is a downward shift of occupied orbitals
compared to empty orbitals with a similar character due to the
self-interaction correction.  Thus, this term correctly opens,
respectively widens, a band gap of transition metal oxides. For this
reason, the inclusion of a scaled Fock term as in the hybrid
functionals is vital for the present work. For the manganites this
affects predominantly the Mn-d orbitals and the Pr-f orbitals.

\begin{figure}[!htbp]
\begin{center}
\includegraphics[width=0.85\linewidth]{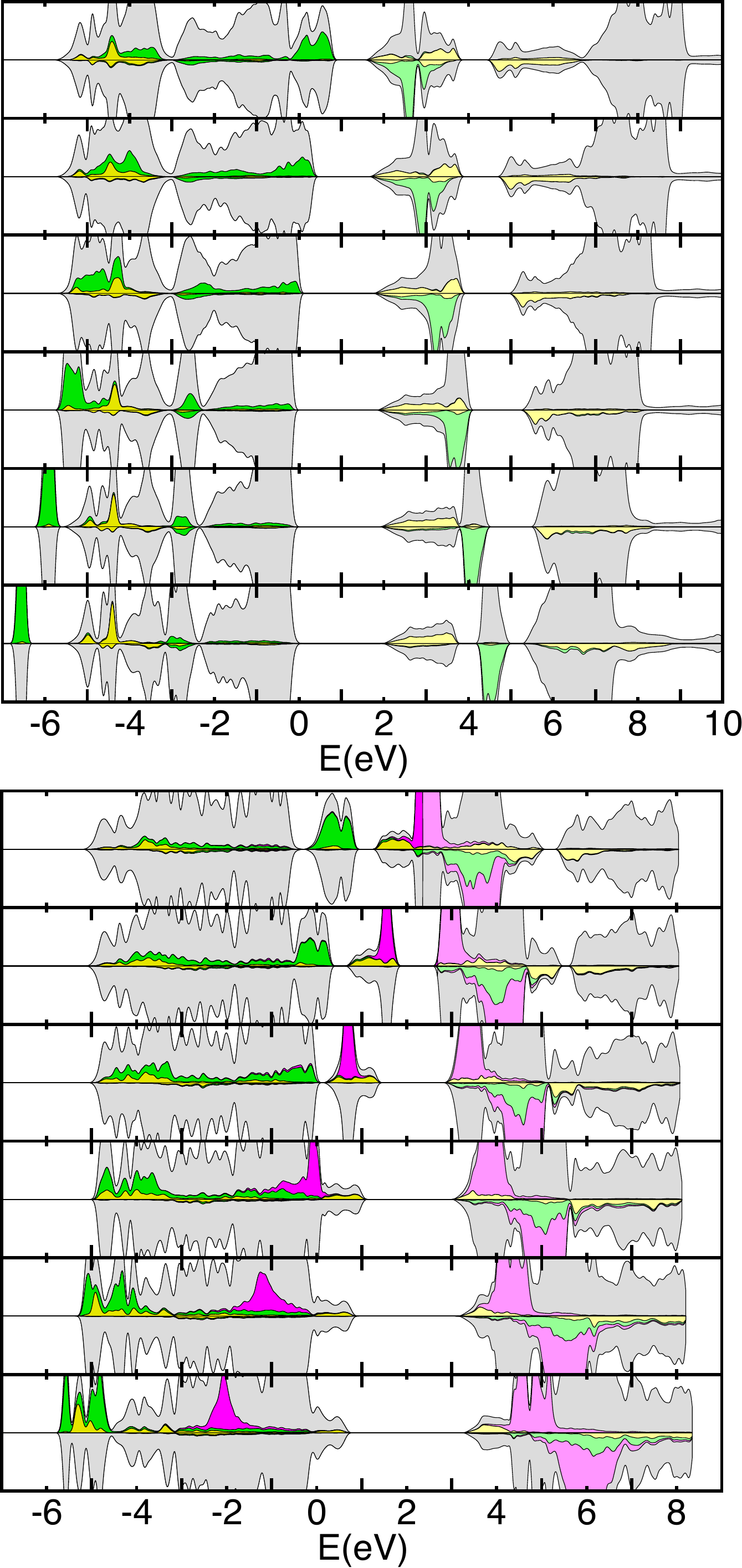}
\end{center}
\caption{\label{fig:utest}DoS for {CaMnO$_3$} (top), and
  {PrMnO$_3$} (bottom) for hybrid mixing factors from 0 to 0.25 in steps
  of 0.05.  to bottom. The hybrid mixing factors are equal on all
  atoms. The energy zero has been set to the top of the O-p band
  for $a_x=0.1$. }
\end{figure}

The effect of the Fock admixture on the DoS of {CaMnO$_3$} and
  {PrMnO$_3$} is shown in figure~\ref{fig:utest}, where the hybrid
  mixing factors have been varied. The hybrid factors on all atoms
  have been kept equal.

The dominant effect on {CaMnO$_3$} is that the filled $t_{2g}$ states
are shifted downward from above the O-p band to its bottom part. A DFT
calculation without admixture of a Fock term, describes the band gap
of {CaMnO$_3$} incorrectly as due to a transition from $t_{2g}$ to
$e_g$ states.  Already a small contribution $a_x=0.05$ of exact
exchange shifts the filled $t_{2g}$ states into the oxygen valence
band and changes the nature of the material from a band
  insulator to a charge-transfer insulator.

A closer look reveals that the $t_{2g}$ states within the O-p valence
band form two distinct contributions,   one due to
  antibonding states in the upper part and another one due to bonding
  states in the lower part of the valence band.

Only for large contributions of the Fock term,
the $t_{2g}$ states leave the oxygen valence band at the bottom and
form a clearly identifiable separate peak. 

A secondary effect of the exact exchange admixture in the hybrid
functionals is the increase of the Hund's-rule coupling between
majority-spin and minority-spin orbitals. The Hund's-rule coupling is
reflected in the separation of the $e_g$ orbitals of both spin
directions. The Hund's-rule coupling grows with increasing
localization of the $t_{2g}$ orbitals, which in turn increases the
local moment.

The position of the oxygen valence band is little affected by the Fock
admixture, despite the fact that the O-p orbitals are filled and
fairly localized.

Turning our attention  to {PrMnO$_3$}  in figure~\ref{fig:utest}, we
observe a dependence of the Mn-d states on the Fock admixture similar
to that in {CaMnO$_3$}. One important difference is that not only the
$t_{2g}$ states, but also the filled $e_g$ orbitals are shifted
downward with increasing the Fock admixture.

A feature in {PrMnO$_3$} not present in {CaMnO$_3$} are the Pr-f
states. For a pure GGA calculation using PBE, the Fermi level is
pinned within the Pr-f states. The Pr-f states are split into
majority- and minority-spin contributions, but there is no splitting
between the occupied and empty states.  The Fock admixture from
  the hybrid functional splits the occupied Pr-f states into separate
  bands of filled and unoccupied states. The filled states are shifted
  downward in energy with increasing Fock admixture and the empty
  states are shifted upward by a similar amount.

\begin{figure}[htbp]
\includegraphics[width=\linewidth]{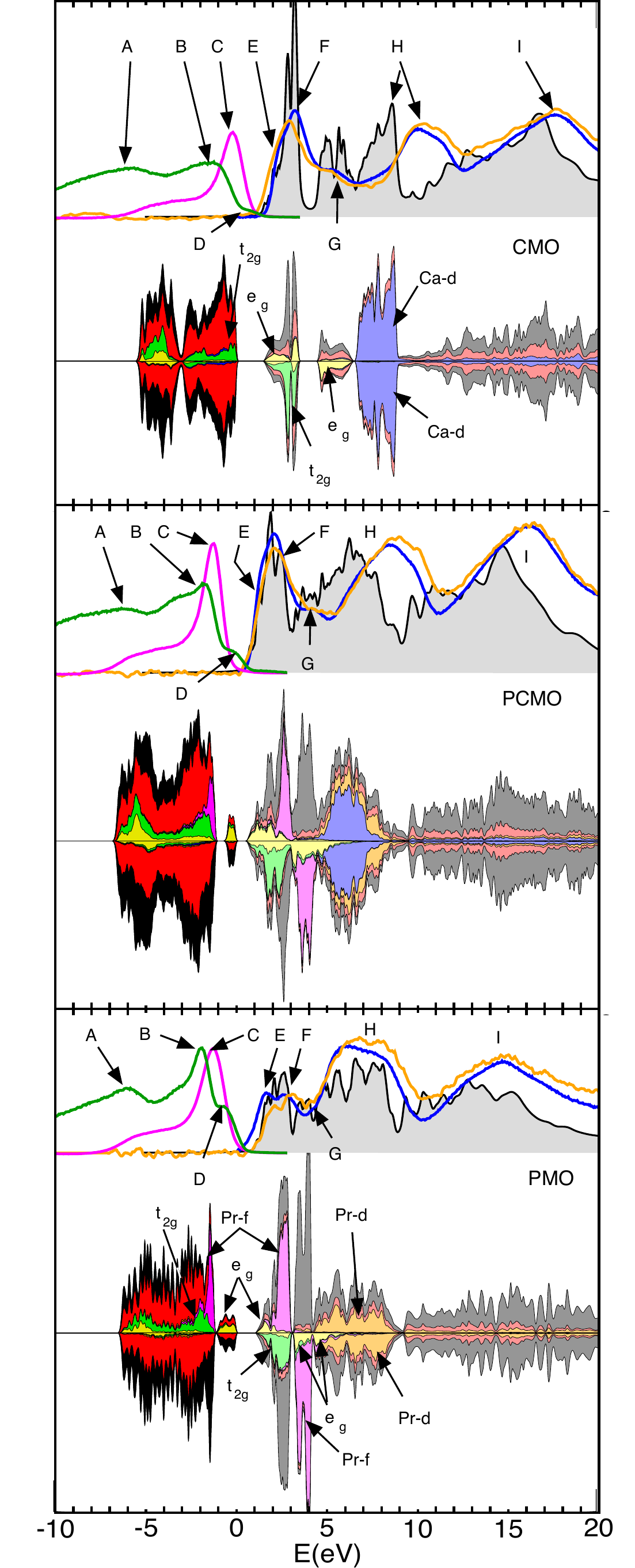}
\caption{\label{fig:pcmodos}Mn-XPS (green), Pr-XPS (magenta), XANES
  (blue), ELNES (orange), together with calculated ELNES spectra
  (shaded grey) and spin-resolved DoS (below). The results are shown
  for {CaMnO$_3$} (top) (x=0.8 for XANES and x=0.95 for ELNES),
  {Pr$_{1/2}$Ca$_{1/2}$MnO$_3$} (middle) and {PrMnO$_3$} (bottom).
  Projected DoS are color coded for Mn-$e_g$ (yellow), Mn-$t_{2g}$
  (green), Pr-f (magenta), Ca-d (blue), Pr-d (orange) and O-p (red).
  The DoS of the two spin directions are shown with opposite
  sign. Only the majority-spin direction is shown for Mn and Pr.}
\end{figure}

\subsection{XPS}
After understanding the role of the Fock admixture, we relate our
findings to experiment and we extract a suitable set of parameters for
the Fock admixture. For this purpose, the XPS data are most
useful. The overall features of non-resonant valence band
photoemission have been discussed by Kurash et
al.\cite{kurash00_matsciengb76_14} Our resonant Mn-XPS, shown as green
line in figure~\ref{fig:pcmodos}, shows a main peak (B) at -2~eV on
our energy scale and a weaker maximum (A) at -6~eV. These two bands
are attributed to the two contributions of the d-orbitals within the
O-p valence band. The broadening of feature (A) towards higher binding
energies is due to the presence of a Mn LVV Auger process, which can
clearly be separated from the resonant Mn 2p$\rightarrow$3d peak as
shown by Wadati et al.\cite{wadati07_jmmm310_963}

Characteristic for the manganites is the shoulder (D) between the
upper peak and the Fermi-level. We attribute this feature with the
filled $e_g$ orbitals located above the O-p band. A clear indication
for this interpretation is the growth of this feature with $1-x$, that
is with increasing number of electrons in the d-shell.  Because the
experimental data are obtained at small but non-zero Pr concentration,
i.e. $x=0.80$, the pre-peak is still visible even at our most Ca-rich
materials.

The dominance of the upper $t_{2g}$ peak (B) compared to the lower one
(A), and the existence of the pre-peak (D) above the main peak in the
Mn-XPS, can only be obtained with narrow range of Fock admixture on
the Mn site.  We have chosen $a_x=0.07$ on the Mn site.

Turning now to the experimental Pr-XPS, a single band (C) from the
occupied Pr-f states is observed which is similar in shape to the
Pr-projected DoS. It is important that the Pr-f band is
located energetically between the main peak of the Mn-XPS and the
pre-peak.  We have chosen a value of $a_x=0.15$ on the Pr site.

The commonly used Fock-admixing factor\cite{perdew96_jcp105_9982} of
$a_x=0.25$ makes the physically realized configuration of f-electron
atoms unstable. As shown above, it furthermore overestimates the
effect on the spectral properties. For this reason we determine the
correction factors by comparison with the experimental spectral
properties.  The smaller Fock admixture is in line with recent
studies.\cite{luo07_prl99_36402,he12_prb86_235117,he12_prb85_195135}
The comparison with experimental spectra suggests that the Fock
admixture should be even smaller than the value $a_x=0.15$ suggested
earlier.\cite{he12_prb86_235117,he12_prb85_195135}

\begin{table}[!ht]
\caption{\label{tab:hybridfactor}Factors $a_x$ for the element
  specific admixture of the Fock term.}
\begin{ruledtabular}
\begin{tabular}{lcccc|}
Element & Pr & Ca & Mn & O\\
\hline
$a_x$ & 0.15 & 0.1 & 0.07 & 0.1 \\
\end{tabular}
\end{ruledtabular}
\end{table}

\subsection{ELNES and XANES}
The O $K$ ELNES and XANES spectra describe the absorption intensity of
an electron transition from the 1s core shell of oxygen to its empty
p orbitals. Therefore, it is sensitive to the antibonding states of
mainly Mn-$e_g$ character. The antibonding $e_g$ states lift some O-p
weight from the valence band into the spectrum of unoccupied states
and are thus probed. The Mn-$t_{2g}$ states contribute less to the
spectrum, because they only form weak antibonds with $\pi$-character.
Another contribution to the spectra is due
to lone pairs of the oxygen bridges, which point towards the A-type
ion. The antibonding interaction with the d-electrons of the A-type
ions, Ca and Pr, lifts some oxygen weight into the empty spectrum
related to the A-type d-states, which is picked up by the spectra.

The experimental ELNES and XANES spectra are shown in
figure~\ref{fig:pcmodos}  alongside with the calculated ELNES spectra
and the calculated DoS. 

The experimental spectra exhibit three main peaks: an upper peak (I)
at 15-18~eV, a middle peak (H) at 6-10~eV and a lower peak (E, F, and G)
below 6~eV.  We assign the broad band (I) near 15-18~eV to a free-electron
like feature.

The middle peak (H) near 6-10~eV is attributed to the Ca-d and Pr-d
orbitals that overlap with the lone-pairs of the oxygen bridges. The
calculated Ca-d and Pr-d states lie a few eV below the
experimental ones. While we can only speculate about the origin at this
point, we attribute this to a self-energy shift, which is not
described adequately in our DFT calculations. For a high-energy
excitation, the electron-correlations in the quasi-particle spectrum,
which reduce the Coulomb repulsion, can not fully develop. In the DFT
calculations in contrast, the correlations are fully developed, so
that the resulting states lie lower than the measured excitation
levels.

The spectral region of interest for the polaron physics of manganites
is related to the lower peak (E, F, G) of the experimental spectra and
lies in the energy window from 1~eV to 6~eV.

The lower peak has a shoulder (G) towards higher energies i.e. at 4-5~eV.
This split-off band at higher energy can be attributed to the
minority-spin Mn-$e_g$ states. It is present for all doping levels.

The main peak (F) of the lower band is due to the majority-spin $e_g$
orbitals and the minority-spin $t_{2g}$ orbitals, which both lie in an
overlapping energy region. Due to their smaller matrix elements, the
$t_{2g}$ states contribute less to the absorption than the $e_g$
states.

For {CaMnO$_3$} the main peak of the lower band in the calculated and
experimental spectra has an interesting structure with a shoulder (E)
towards lower energies.  Also, the calculated DoS of the
majority-spin $e_g$ states in that energy region can be divided into a
sharp feature at the upper band edge attributed to feature (F) and a
broad tail extending towards lower energies attributed to feature (E).

Proceeding from {CaMnO$_3$} towards {PrMnO$_3$}, the relative weight
of the bands (E, F) associated with the majority-spin Mn-$e_g$ states
at 1-4~eV decreases. The weight is measured relative to the
high-energy bands (H,I) due to the A-type d electrons and the free
electron feature.  This decrease in weight can be attributed to the
filling of the two majority-spin Mn-$e_g$ levels. Only the empty
orbitals are visible in the ELNES and XANES spectra. From {CaMnO$_3$}
to {PrMnO$_3$} the number of empty majority-spin orbitals is reduced
by a factor of two, which is consistent with the corresponding
decrease of weight.

The experimental ELNES and XANES spectra of {PrMnO$_3$} exhibit a double
peak (E, F) in the energy region 1-3~eV. The calculated spectra exhibit
one broad band in this energy window. However, two fairly different
states contribute: in the lower part (E) the majority-spin $e_g$
states are dominant, while in the upper part (F) the minority-spin
$t_{2g}$ states contribute. Therefore, we attribute the double peak to
these two contributions.

\section{Model Hamiltonian}
\label{sec:model}

In this section, we present a simple model that describes the relevant
physical effects.  The physics of manganites is characterized by
strong correlations of electrons, phonons and classical spins. The
electrons are those in the two $e_g$ orbitals of the Mn ions. The
three spin-aligned electrons in the $t_{2g}$ states of the Mn ions are
described as classical spins.  The two Jahn-Teller active distortions
and an isotropic expansion of the MnO$_6$ octahedra are the relevant
phonons, which are strongly coupled to the electrons.  All other
degrees of freedom are either absorbed into the dynamical variables of
the model, or they are considered as a bath and not treated
explicitly.

A variety of models of this kind have been investigated and we refer
to the excellent review of Hotta.\cite{hotta06_rpp69_2061} The
selection of the model and the choice of the model parameters are
obtained either to match experimental observations or alternatively
from first-principles calculations.  While the first way is
insensitive to the errors inherent in ab-initio calculations, it is
possible that the same set of observations can be produced by
different sets of parameters. This obscures the connection between
the underlying physics and the experimental facts. Proceeding from
ab-initio calculations, we hope to provide a consistent set of parameters
that correctly captures the underlying physics.

\subsection{Functional form}
In the model, the total-energy functional of density-functional theory
is replaced by a potential energy functional of the form
\begin{equation}
\label{equ:pot-energy-fctal}
E_{pot} = E_{e}+E_{S}+E_{ph}+ E_{e-ph} +E_{e-S}\;, 
\end{equation} 
where $E_e$ is the energy of the isolated electronic subsystem.  $E_S$
describes the spin system, namely the antiferromagnetic interaction of
the $t_{2g}$ orbitals on neighboring Mn sites. $E_{ph}$ is the phonon
energy. The remaining two terms describe the coupling of these
subsystems, namely the electron-phonon coupling $E_{e-ph}$ due to the
Jahn-Teller effect and the Hund's coupling $E_{e-S}$ between
electrons in the $e_g$ and $t_{2g}$ shells of a given Mn site.

The electrons are described, in the spirit of density-functional
theory\cite{hohenberg64_pr136_B864,kohn65_pr140_1133} by a set of
one-particle wave functions.  In order to account for non-collinear
magnetic structures, each wave function is a two-component spinor.
The one-particle wave function with band index $n$ is expressed as
\begin{equation}
|\psi_n\rangle=\sum_{\sigma,\alpha,R}|\chi_{\sigma,\alpha,R}\rangle
\psi_{\sigma,\alpha,R,n}
\end{equation} 
in terms of local spin-orbitals $|\chi_{\sigma,\alpha,R}\rangle$
having spin $\sigma\in\{\uparrow,\downarrow\}$, spatial orbital
character $\alpha\in\{a,b\}$ denoting the $d_{x^2-y^2}$ orbital for
$\alpha=a$ and $d_{3z^2-r^2}$ for $\alpha=b$ orbitals. The orbital is
centered at the Mn site specified by the index $R$. The orbitals are
aligned with the local coordinate axes defined by octahedral axes.

The electronic energy $E_e=E_{kin}+E_U$ consists of the kinetic energy
$E_{kin}$ and the electron-electron interaction $E_U$.

The kinetic energy of the $e_g$ electrons is  
\begin{equation}
 E_{kin}=\sum_{R,R'} \sum_{n} f_n \sum_\sigma\sum_{\alpha,\beta}
 \psi^*_{\sigma,\alpha,R,n}
 T_{\alpha,\beta,R,R'}\psi_{\sigma,\beta,R',n} \;.
\label{eq:E-kin}	 		 
\end{equation}
The hopping-matrix elements contribute only onsite and
nearest-neighbor terms between the Mn sites.

The overall form of the hopping matrix is obtained assuming orbitals
with pure angular momentum character. It is obtained by down-folding
the p-orbitals of the oxygen bridge to obtain the indirect hopping
matrix element between the d-orbitals with $e_g$ character. Only
the axially symmetric orbitals along the bond Mn-O axis can
contribute. Due to its symmetry, the d-orbital with $\delta$ character
along the bond axis does not couple to a p-orbital of the bridging
oxygen ion.

For nearest neighbors along the $\pm z$-direction, we obtain
\begin{eqnarray}
{\bm T}_{R,R'}&=&-t_{hop}
\left(\begin{array}{cc}0 & 0 \\0 & 1 \end{array}\right)
\;,
\end{eqnarray}
while the hopping matrix elements in the $\pm x$-direction is
\begin{eqnarray}
{\bm T}_{R,R'}&=&-t_{hop}
\left(\begin{array}{cc}3/4 & -\sqrt{3}/4 \\-\sqrt{3}/4 & 1/4 \end{array}\right)
\end{eqnarray}
and the one along the $\pm y$-direction is
\begin{eqnarray}
{\bm T}_{R,R'}&=&-t_{hop}
\left(\begin{array}{cc}3/4 & \sqrt{3}/4 \\\sqrt{3}/4 & 1/4 \end{array}\right)
\;.
\end{eqnarray}

In addition to the intersite matrix elements, the hopping contributes
the onsite matrix.
\begin{eqnarray}
{\bm T}_{R,R}&=&t_{hop}
\left(\begin{array}{cc}3 & 0 \\ 0 & 3 \end{array}\right)
\;.
\end{eqnarray}
The onsite term ensures that a superposition of two d-orbitals on
either side of an oxygen bridge cannot profit from bonding if their
parity differs from that of the bridging O-p orbitals. 

The down-folding produces also a constant energy term $-6t_{hop}$ for
each Mn site, which accounts for the Mn-O bonding contribution. We
omit it, because it is a global constant. However, this term may
become relevant when site-dependent and material-specific hopping
parameters are used.

The Coulomb energy for a diagonal density matrix will be expressed by
the local one-center reduced density matrix
\begin{eqnarray}
\rho_{\sigma,\alpha,\sigma',\beta,R}=\sum_n
f_n\psi_{\sigma,\alpha,R,n}\psi^*_{\sigma',\beta,R,n}
\;.
\end{eqnarray}

The onsite Coulomb energy between the $e_g$ electrons
\begin{eqnarray}
\label{eq:E-Col}
E_U=E_H+E_{SIC}+E_{OR}
\end{eqnarray}
can be broken up into a term proportional to the Hartree energy
\begin{eqnarray}
\label{eq:E-hartree}
E_{H}&=&
\frac{1}{2}(U-3J_{xc})\sum_{R}
\left(\sum_{\sigma,\alpha}
\rho_{\sigma,\alpha,\sigma,\alpha,R}\right)^2
\end{eqnarray}
and a corresponding self-interaction correction
\begin{eqnarray}
\label{eq:E-SIC}
E_{SIC}&=&-\frac{1}{2}(U-3J_{xc})\sum_R
\sum_{\sigma,\alpha,\sigma',\beta}
|\rho_{\sigma,\alpha,\sigma',\beta,R}|^2
\;.
\end{eqnarray}
The self-interaction term stabilizes
filled orbitals, and is thus largely responsible for opening up the
band gap in transition metal oxides with a partially filled
d-shell. 

The remaining contribution $E_{OR}$ to the onsite Coulomb interaction is
\begin{eqnarray}
\label{eq:E-OP}
E_{OR}
&=&\frac{1}{2}J_{xc}\sum_R\sum_{\sigma,\sigma'}(-1)^{\sigma-\sigma'}\sum_{k\in\{x,z\}}
\nonumber\\
&\times&\Bigl[
\Bigl(\sum_{\alpha,\beta}\rho_{\sigma,\alpha,\sigma',\beta,R}
\sigma^{(k)}_{\beta\alpha}\Bigr)
\Bigl(\sum_{\alpha,\beta}\rho_{-\sigma,\alpha,-\sigma',\beta,R}
\sigma^{(k)}_{\beta\alpha}\Bigr)
\nonumber\\
&&
+\Bigl(\sum_{\alpha}\rho_{\sigma,\alpha,\sigma',\alpha,R}\Bigr)
\Bigl(\sum_{\alpha}\rho_{-\sigma,\alpha,-\sigma',\alpha,R}\Bigr)\Bigr]
\label{eq:eor}
\end{eqnarray}
With $\sigma^{(k)}_{\alpha,\beta}$ we denote the three Pauli matrices
for $k\in\{x,y,z\}$.  The notation $-\sigma$ implies
$-\sigma=\uparrow$ for $\sigma=\downarrow$ and vice versa. Similarly,
$(-1)^{\sigma-\sigma'}=1$ for $\sigma=\sigma'$ and
$(-1)^{\sigma-\sigma'}=-1$ for $\sigma\neq\sigma'$.  The term $E_{OR}$
vanishes for a fully spin-polarized system.

The division of the Coulomb energy used here is advantageous for the
manganite systems. In manganites, the $e_g$ shell is
nearly fully spin polarized so that $E_{OR}$ is small. It has little
effect on the filled orbitals, while it affects the position of the
minority-spin states.  The decomposition also makes it evident that
the effective Coulomb parameter consists of a combination of several
Kanamori parameters, $U$ and $J_{xc}$, which tend to compensate each
other to a large extent.

The phonons are described by classical amplitudes $Q_{1,R}$ of
the breathing mode, the uniform octahedral expansion, and those,
$Q_{2,R}$ and $Q_{3,R}$, of the two Jahn-Teller active phonon
modes.\cite{kanamori60_japplphys31_S14} They are defined by
\begin{eqnarray}
Q_{1,R}&=&\frac{1}{\sqrt{3}}(d_{x,R}+d_{y,R}+d_{y,R}-3\bar{d})
\nonumber\\
Q_{2,R}&=&\frac{1}{\sqrt{2}}(d_{x,R}-d_{y,R})
\nonumber\\
Q_{3,R}&=&\frac{1}{\sqrt{6}}(-d_{x,R}-d_{y,R}+2d_{z,R})\;,
\end{eqnarray}
where $d_{x,R}$ is the distance between the oxygen atoms on the left
and right corners of the octahedron centered at $R$. Similarly $d_y$
is the distance of the oxygen atoms in the front and back and
$d_{z,R}$ is the distance of the oxygen atoms on the top and bottom
corners. $\bar{d}$ is the equilibrium distance of two opposite oxygen
atoms in the octahedron. The $Q_2$ mode specifies the compression and
expansion along orthogonal axes in the $xy$-plane.  The $Q_3$ mode
describes expansion in $z$-direction and compression in the $xy$
plane. 

The electron-phonon coupling $E_{e-ph}$ is
\begin{equation}
\label{eq:E-vib-04-08-2016}
E_{e-ph} = g_{JT} 
\sum_{R,\sigma}\sum_{\alpha,\beta} \rho_{\sigma,\alpha,\sigma,\beta,R}
M^Q_{\beta,\alpha}(Q_{1,R},Q_{2,R},Q_{3,R}) \,,
\end{equation}
where $g_{JT}$ and $g_{br}$ are the electron-phonon coupling constants
and
\begin{eqnarray}
\mathbf{M}^Q(Q_1,Q_2,Q_3)=
\left(\begin{array}{cc}
Q_{3} & Q_{2}\\Q_{2} & -Q_{3}
\end{array}\right)-{\bm 1}\frac{g_{br}}{g_{JT}}Q_1\;.
\end{eqnarray}

The phonon energy $E_{ph}$ describes the term restoring the symmetric
octahedron
\begin{equation}
E_{ph}=\frac{1}{2}k_{JT}\sum_{R}\left(Q_{2,R} ^2+Q_{3,R}^2 \right)
+\frac{1}{2}k_{br}\sum_R Q_{1,R}^2
\;,
\end{equation}
where $k_{JT}$ is the restoring force constant for the Jahn-Teller
distortions and $k_{br}$ is the restoring force constant of the
breathing distortion.

We describe the three majority-spin $t_{2g}$ electrons at site $R$ by
their classical spin $\vec{S}_R$. While the direction of the spin may
vary, the magnitude of the spin vector is fixed to
$|\vec{S}_R|=\frac{3}{2}\hbar$.

The spin energy 
\begin{eqnarray}
E_S=\frac{1}{2}J_{AF}
\sum_{R,R'}\delta_{|\vec{R}-\vec{R'}|-1}
\left(\frac{3\hbar}{2}\right)^{-2}
\vec{S}_R\vec{S}_{R'}
\end{eqnarray}
 is due to a small anti-ferromagnetic coupling of the spins on
 neighboring sites. 

The spins $\vec{S}_R$ of the $t_{2g}$ electrons are strongly coupled to
the spins of the $e_g$ electrons by the Hund's coupling $J_H$.
The Hund's coupling is described by
\begin{equation}
\label{eq:E-Hund}
E_{e-S} = -J_H
\sum_{R,\alpha}
\sum_{\sigma,\sigma'} 
\rho_{\sigma,\alpha,\sigma',\alpha,R}
M^S_{\sigma',\sigma}(\vec{S}_R)
\end{equation}
where
\begin{eqnarray}
 \mathbf{M}^S(\vec{S})=\left(\frac{3\hbar}{2}\right)^{-1}\left(
 \begin{array}{cc}
 S_{z} & S_{x}- iS_{y}  \\
 S_{x}+ iS_{y}  & -S_{z} 
 \end{array} \right)
\;.
\end{eqnarray}

\subsection{Parameter determination}
\label{sec:MD-details}
The ratio of the breathing and Jahn-Teller parameters have been
determined from theoretical grounds: because the Jahn-Teller splitting
is due to Mn-O antibonds, an axial elongation affects only the
position of the axial d-orbital, while leaving the d-orbital with
$\delta$-symmetry about the axis unchanged. A pure elongation along a
single axis has a fixed ratio of Jahn-Teller distortion and 
breathing amplitude, namely 
\begin{eqnarray}
\sqrt{Q_2^2+Q_3^2}=Q_1\sqrt{2}
\;.
\label{eq:ratioQ123}
\end{eqnarray}

Such a distortion leaves one level unchanged, when the ratio of the
coupling constants is $g_{br}=\sqrt{2}g_{JT}$. We use this condition
to link $g_{JT}$ and $g_{br}$. The restoring force constant $k_{br}$
for the breathing mode is determined such that the equilibrium
distortion of an isolated octahedron due to occupying a $d_{3z^2-r^2}$
orbital is a pure axial elongation, which requires
$g_{br}/k_{br}=(g_{JT}/k_{JT})/\sqrt{2}$, hence $k_{br}=2k_{JT}$.
Thus, we use
\begin{eqnarray}
g_{br}&=&g_{JT}\sqrt{2}
\label{eq:gbrandgjt}
\\
k_{br}&=&2k_{JT}\;.
\label{eq:kbrandkjt}
\end{eqnarray}

The other parameters of the model have been determined from the
first-principles calculations of {CaMnO$_3$} and {PrMnO$_3$} in the stable
magnetic structures.  For this purpose, we selected a set of
quantities, which are, on the one hand, physically relevant and, on
the other hand, expressed by only a small set of parameters at a time.

Many of the relevant energy scales are defined already by the onsite
contribution of the energy functional. In the onsite approximation of
the model, this implies that we restrict the density matrix to a
diagonal form. 
\begin{eqnarray}
\rho_{\sigma,\alpha,\sigma',\beta,R}=n_{\sigma,\alpha,R}
\delta_{\sigma,\sigma'}\delta_{\alpha,\beta}
\end{eqnarray}
Furthermore, we exploit isotropy of the onsite approximation in the
$(Q_2,Q_3)$-plane and in spin space: We restrict the Jahn-Teller
distortion to the $Q_3$ contribution, and we restrict the classical
spin $\vec{S}_R$ to point into the positive $z$-direction.

This model can be optimized with respect to the phonon amplitudes
$Q_1$ and $Q_3$, which expresses them as function of occupations
$n_{\sigma,\alpha}$.
\begin{eqnarray}
Q_3=\frac{g_{JT}}{k_{JT}}\left(\sum_{\sigma}n_{\sigma,b}-n_{\sigma,a}\right)
\nonumber\\
Q_1=\frac{g_{br}}{k_{br}}\left(\sum_{\sigma}n_{\sigma,a}+n_{\sigma,b}\right)
\label{eq:optjtdist}
\end{eqnarray}

The onsite total energy for optimized phonon amplitudes has the form
\begin{eqnarray}
E[n_{\sigma,\alpha}]&=&-J_H
\sum_\alpha (n_{\uparrow,\alpha}-n_{\downarrow,\alpha})
\nonumber\\
&-&\frac{g^2_{JT}}{2k_{JT}}2\sum_\alpha \Bigl(\sum_\sigma n_{\sigma,\alpha}\Bigr)^2
\nonumber\\
&+&\frac{1}{2}(U-3J_{xc})
\Bigl[\Bigl(\sum_{\alpha,\sigma}n_{\alpha,\sigma}\Bigr)^2
-\sum_{\alpha,\sigma}n_{\alpha,\sigma}^2\Bigr]
\nonumber\\
&+&J_{xc}\Bigl[2\sum_\alpha n_{\uparrow,\alpha}n_{\downarrow,\alpha}
+\Bigl(\sum_\alpha n_{\uparrow,\alpha}\Bigr)
\Bigl(\sum_\alpha n_{\downarrow,\alpha}\Bigr)\Bigr]
\nonumber\\
\label{eq:onsitemodelenergy}
\end{eqnarray}
where the parameters for the breathing have been expressed by those for
the Jahn-Teller distortion as described in Eqs.~\ref{eq:gbrandgjt} and
\ref{eq:kbrandkjt}.

We determine the energy levels from the model using Janak's
theorem\cite{janak78_prb18_7165} as derivative of the total energy
with respect to the occupations
$\epsilon_{\sigma,\alpha}=\frac{\partial E}{\partial
  n_{\sigma,\alpha}}$. Finally, we choose a special set of
occupations, namely
\begin{eqnarray}
\rho_{\uparrow,a,\uparrow,a}=n_a;\qquad
\rho_{\uparrow,b,\uparrow,b}=n_b
\end{eqnarray}
while the other occupations, namely
$\rho_{\downarrow,\alpha,\downarrow,\alpha}$ for $\alpha\in\{a,b\}$,
are set to zero.

The resulting energy levels are 
\begin{eqnarray}
\left(\begin{array}{c}
\epsilon_{a,\uparrow}\\
\epsilon_{b,\uparrow}\\
\epsilon_{a,\downarrow}\\
\epsilon_{b,\downarrow}
\end{array}\right)
&=&J_H
\left(\begin{array}{r}
-1\\-1\\1\\1
\end{array}\right)
-2\frac{g_{JT}^2}{k_{JT}}
\left(\begin{array}{c}
n_a\\n_b\\n_a\\n_b
\end{array}\right)
\nonumber\\
&&\hspace{-2cm}
+(U-3J_{xc})
\left(\begin{array}{c}
n_b\\n_a\\n_a+n_b\\n_a+n_b\\
\end{array}\right)
+J_{xc}
\left(\begin{array}{c}
0\\0\\3n_a+n_b\\n_a+3n_b\\
\end{array}\right)
\label{eq:onsiteenergylevels}
\end{eqnarray}
The comparison of these energy levels with the DoS of
our first-principles calculations provides conditions that determines
the values of the model parameters. The details on the relation
between DoS and Hamilton matrix elements are provided in
appendix~\ref{app:hfromdos}.

The parameters have been extracted from {PrMnO$_3$}, which we represent in
the model calculations by $n_b=1$ and $n_a=0$, and {CaMnO$_3$} for which
$n_a=n_b=0$.
\begin{enumerate}
\item From the spin splitting $\Delta_H(x=1)$ of the $e_g$ orbitals in
  {CaMnO$_3$} one extracts the Hund's-coupling parameter
  $\Delta_H(x=1)=2J_H$.
\item The Jahn-Teller splitting $\Delta^{\uparrow}(x=0)$ of the majority-spin
  states in {PrMnO$_3$}   yields
  $\Delta^{\uparrow}(x=0)=2\frac{g_{JT}^2}{k_{JT}}+U-3J_{xc}$
\item The Jahn-Teller splitting $\Delta^\downarrow$ of the
  minority-spin states in {PrMnO$_3$} yields
  $\Delta^{\downarrow}(x=0)=2\frac{g_{JT}^2}{k_{JT}}-2J_{xc}$.
\item From the splitting $\Delta_H(x=0)$ between majority- and
  minority-spin levels in {PrMnO$_3$}, we obtain
  $\Delta_H(x=0)=\Delta_H(x=1)+(U+J)/2$.
\item Using the calculated Jahn-Teller distortion in {PrMnO$_3$}, we can
  extract the electron-phonon coupling constant $g_{JT}$ and the
  restoring force constant $k_{JT}$ separately.
\item Using the Jahn-Teller distortions of {PrMnO$_3$} in the A-type
  magnetic structure, we disentangle $g_{JT}$ and $k_{JT}$ via
  $\sqrt{Q_2^2+Q_3^2}=\frac{g_{JT}}{k_{JT}}$.
\end{enumerate}

\begin{table}[!htb]
 \caption{\label{tab:gcabece}Spin energy $E_S$ per Mn site in units of
   $J_{AF}$ for the magnetic orders according to the notation of
   Wollan.\cite{wollan55_pr100_545} The sketches below show cutouts of
   the spin arrangements. White and black spheres represent Mn sites
   with the majority-spin direction up and down, respectively.}
\begin{center}
  {\setlength{\tabcolsep}{10pt}
  \renewcommand{\arraystretch}{1.2}
  \begin{tabular}
{>{\centering\arraybackslash} m{0.9cm} >{\centering\arraybackslash} m{1.5cm} 
 >{\centering\arraybackslash} m{2cm}}
  \hline
  \hline
  Type & $E_S/J_{AF}/$Mn & Magnetic configuration \\
  \hline
  G(AFM)  & -3  & 
\includegraphics[width=0.7\linewidth]{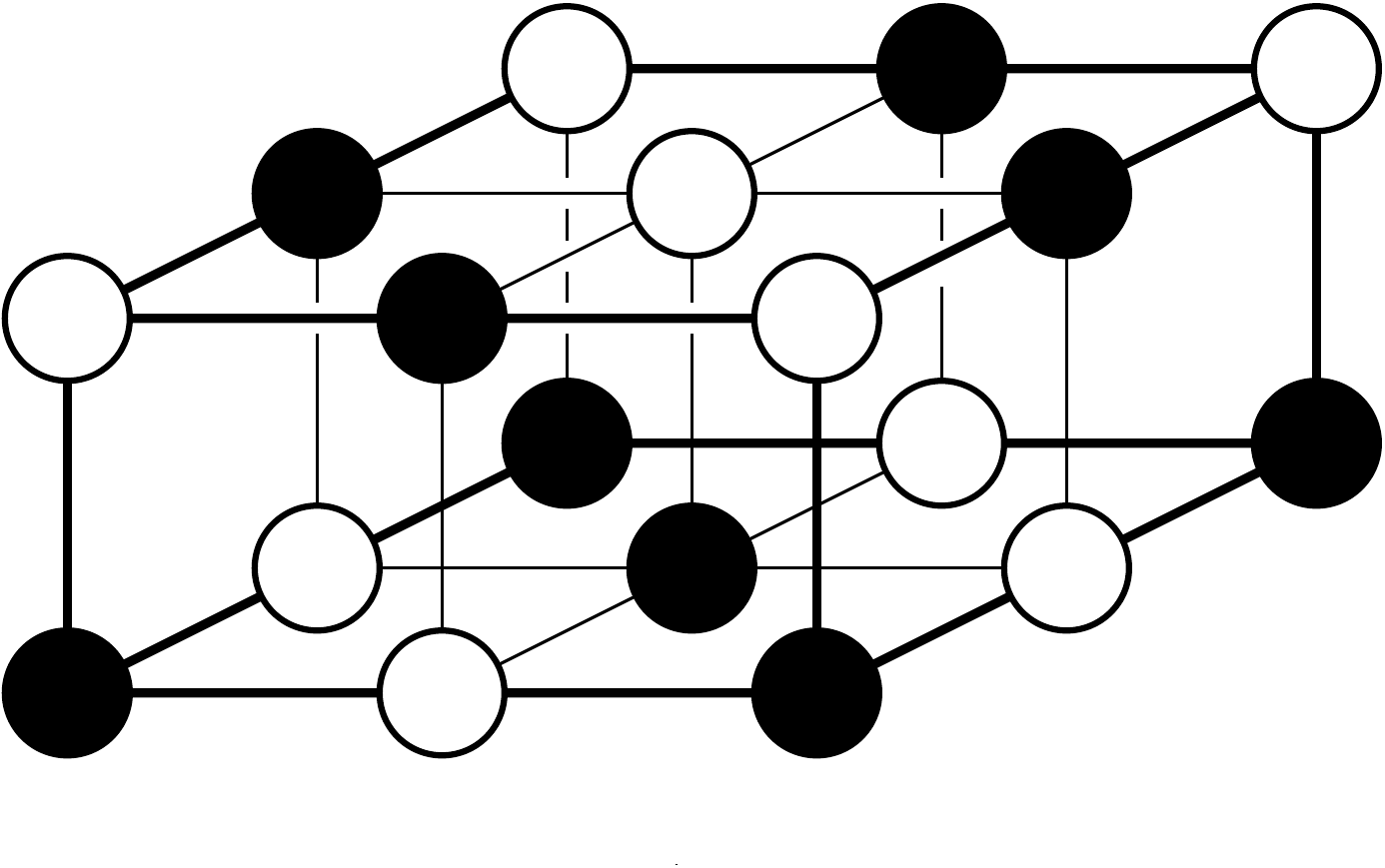}   \\
  C(AFM)  & -1  & 
\includegraphics[width=0.7\linewidth]{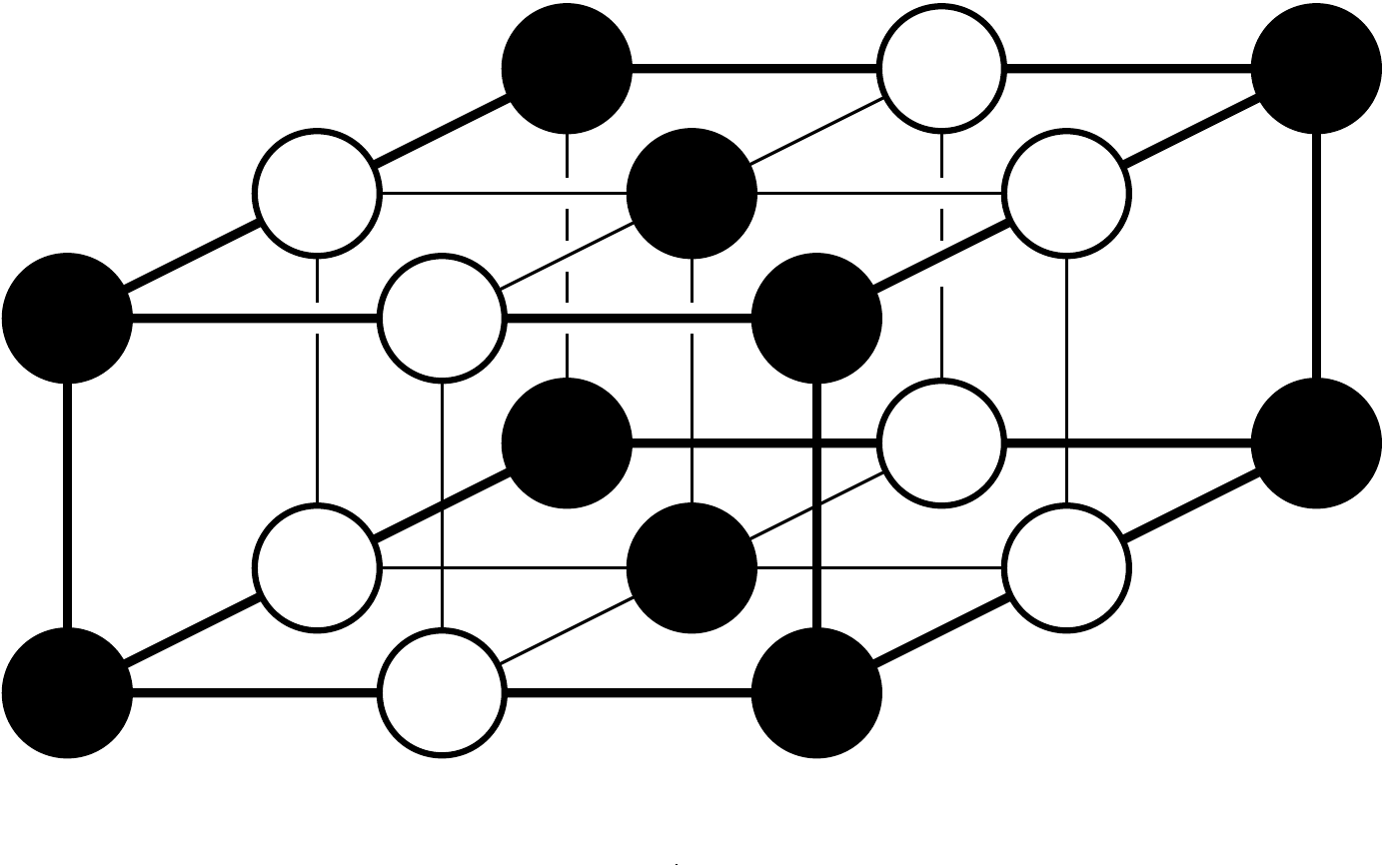}   \\
  CE(AFM) & -1  & 
\includegraphics[width=0.7\linewidth]{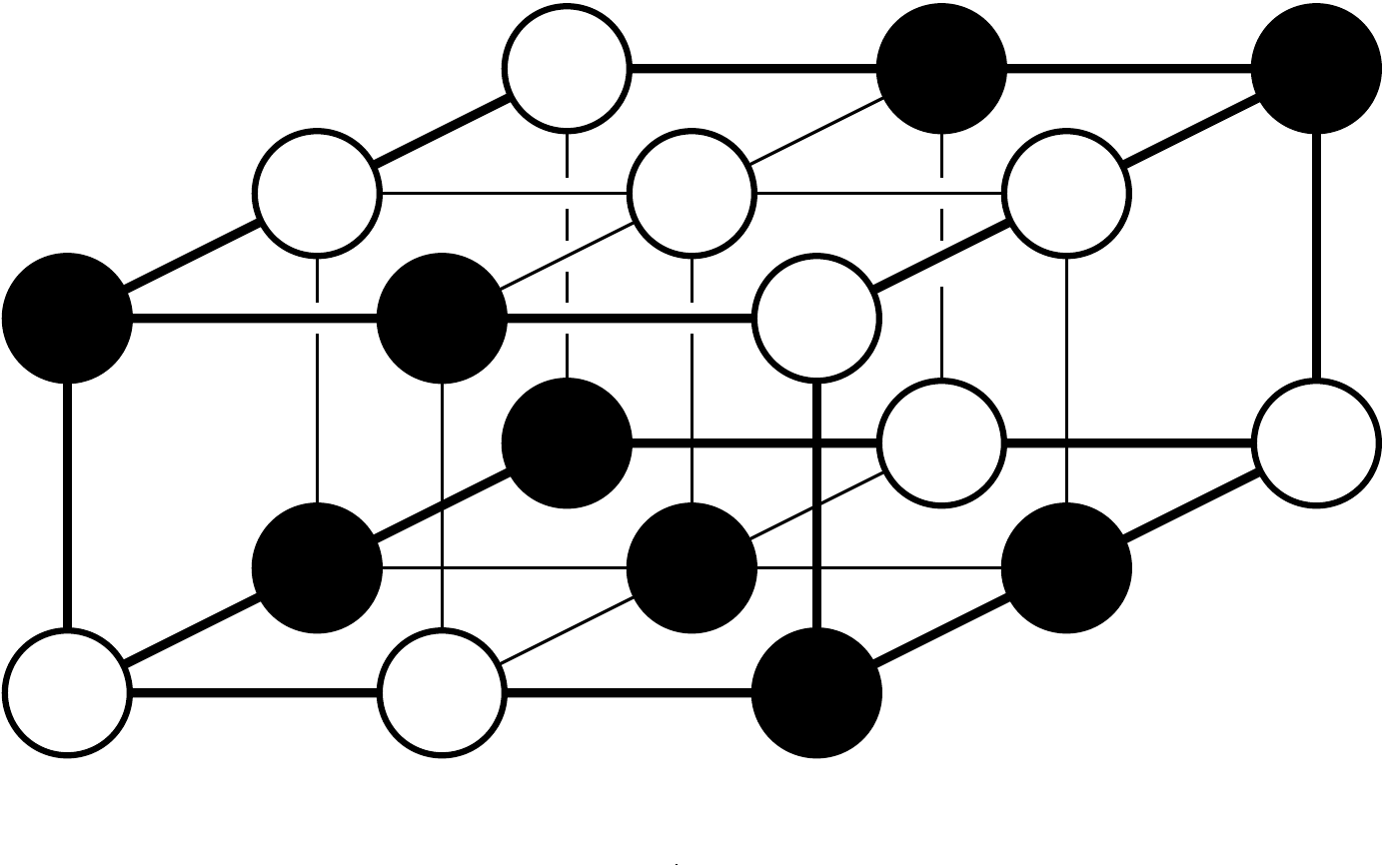}  \\
  E(AFM)  & -1  & 
\includegraphics[width=0.7\linewidth]{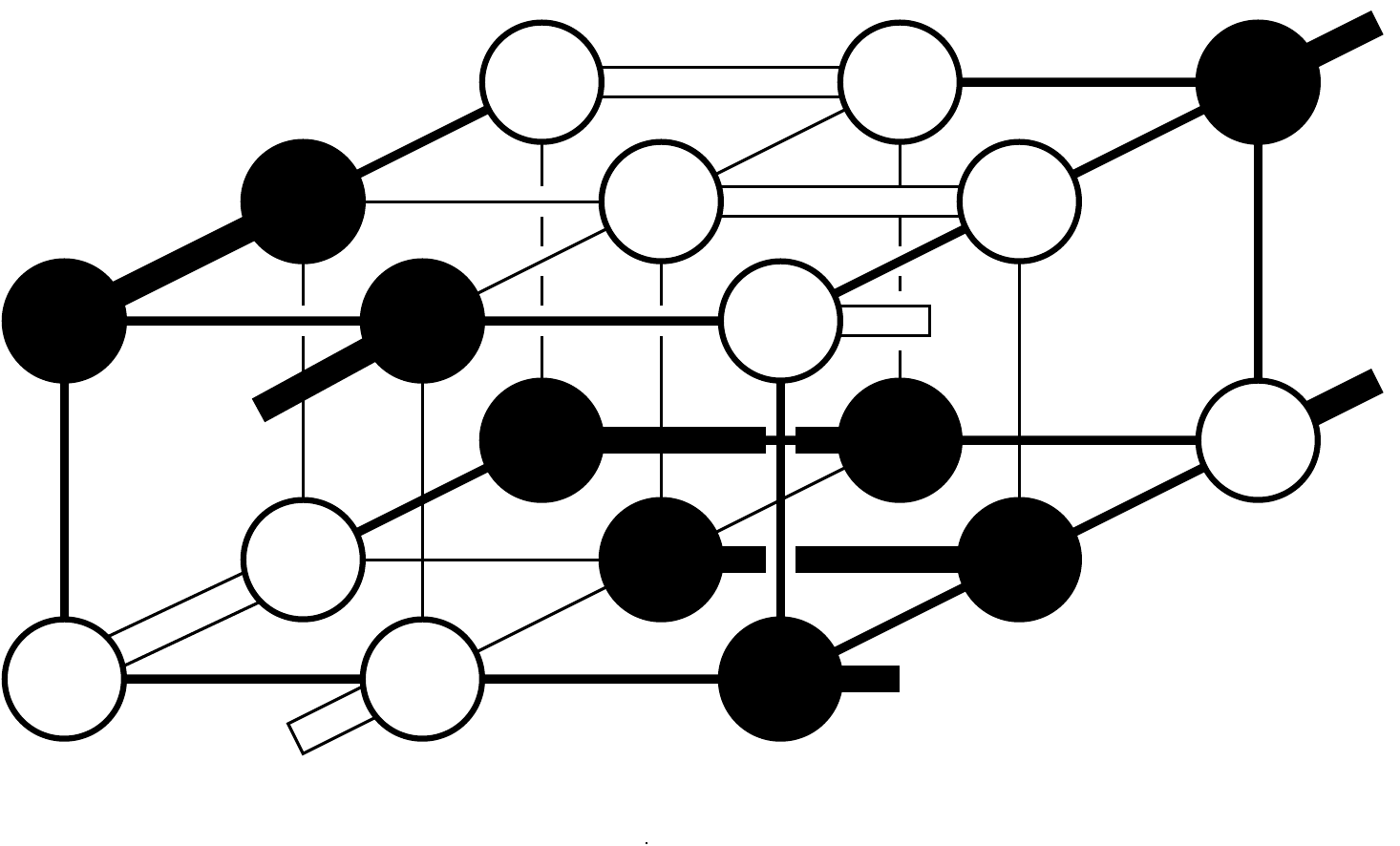}   \\
  A(AFM)  &  1  & 
\includegraphics[width=0.7\linewidth]{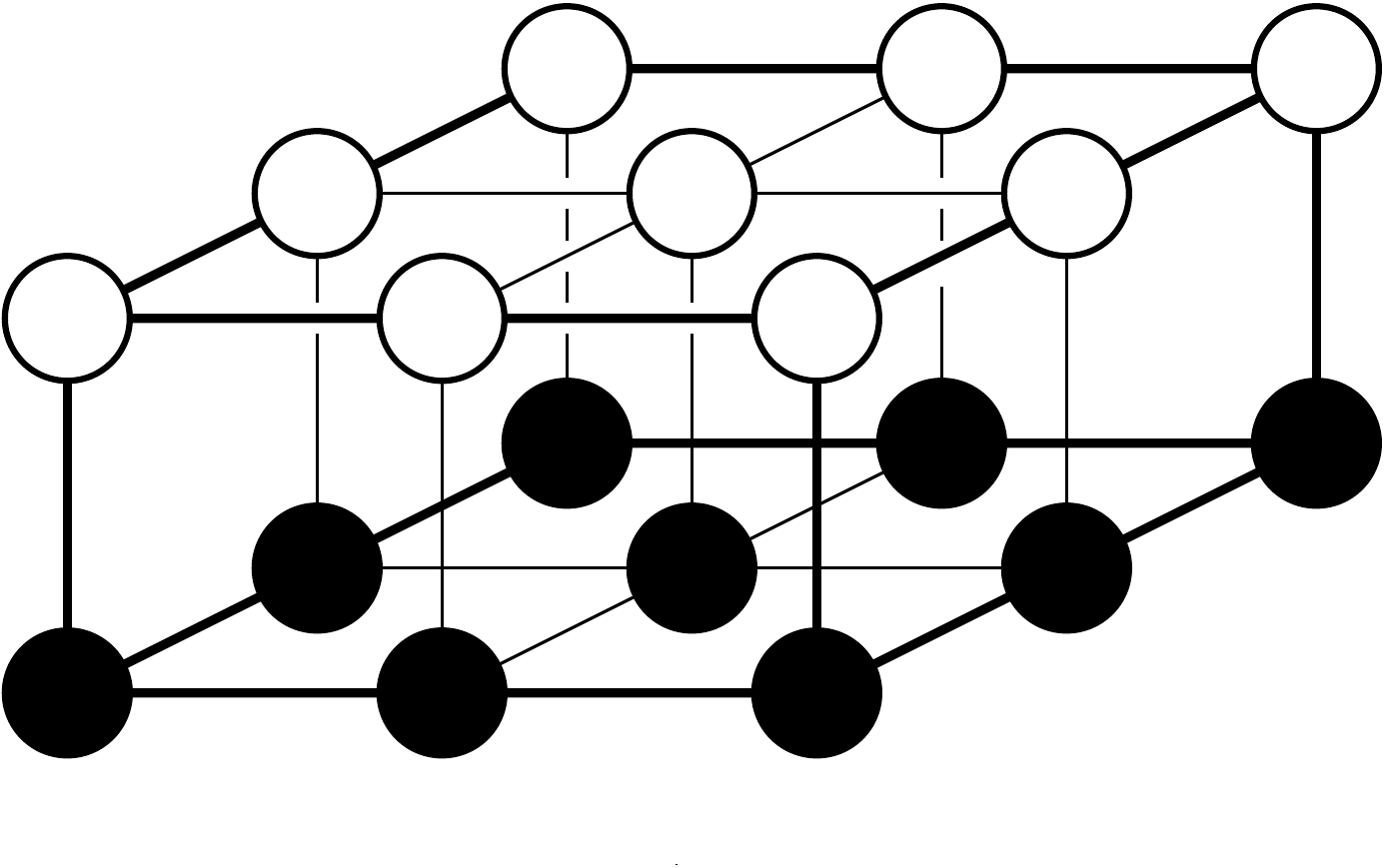}   \\
  B(FM)   &  3  & 
\includegraphics[width=0.7\linewidth]{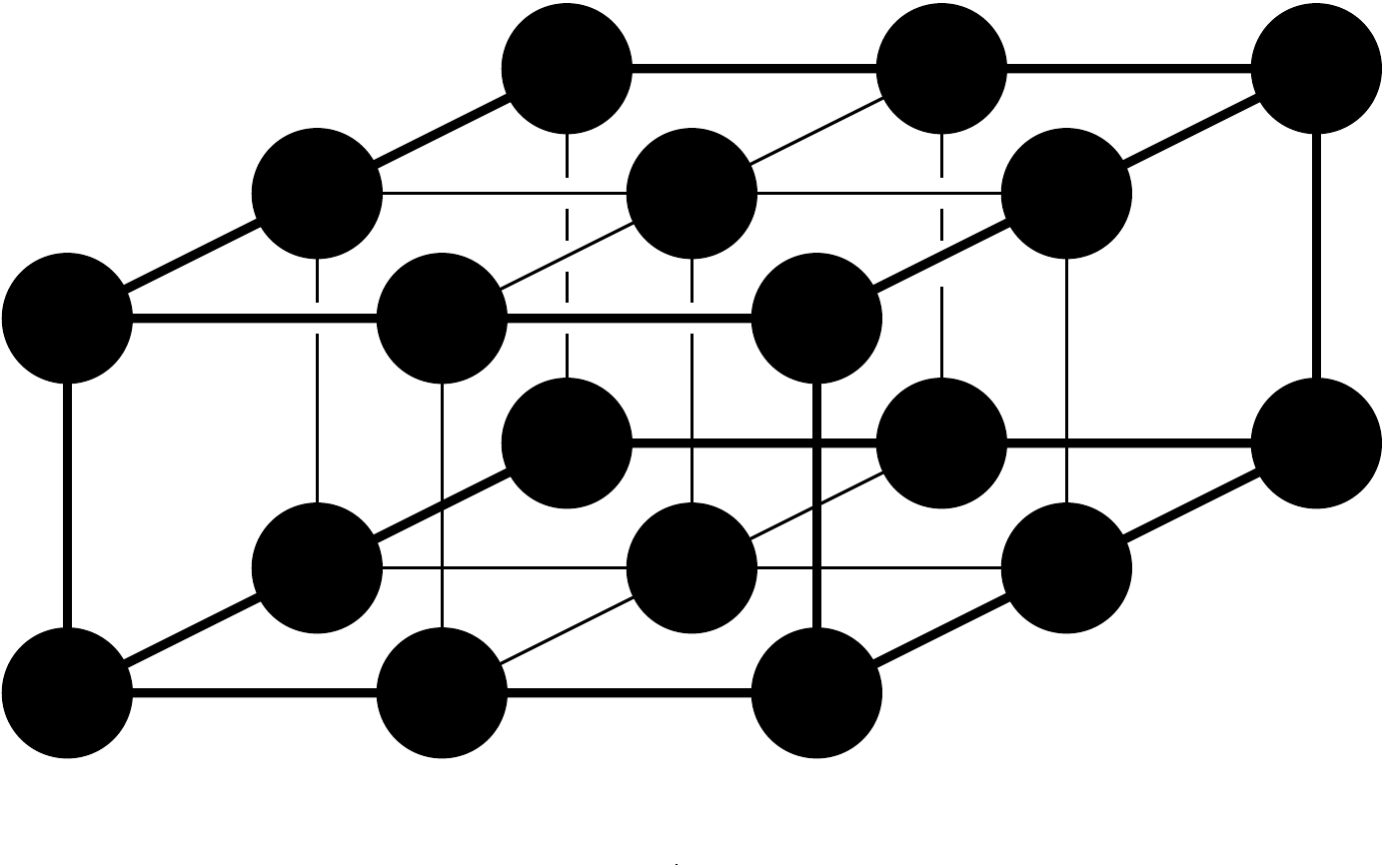}   \\
  \hline
  \hline
  \end{tabular}}\\
\end{center}
\end{table}

The energy $E_S$ related to the Heisenberg coupling can be
expressed for the collinear magnetic structure from the number of
ferromagnetic versus antiferromagnetic Mn neighbors.  The result is
summarized in table~\ref{tab:gcabece}.

From the total energy difference between {CaMnO$_3$} for the G-type
antiferromagnetic order and the ferromagnetically ordered material, we
obtain the value of $J_{AF}=3.326$~meV. With this value the model
predicts the incorrect ground state for the half-doped material,
namely, the ferromagnetic rather than the CE-type antiferromagnetic
order. In order to recover the correct energetic order we increased
the antiferromagnetic coupling to 14~meV.

The hopping parameter $t_{hop}=0.585~eV$ has been extracted as an
average over all oxygen bridges, dopings and magnetic structures
investigated.  

\begin{figure}[!ht]
\begin{center}
\includegraphics[width=\linewidth]{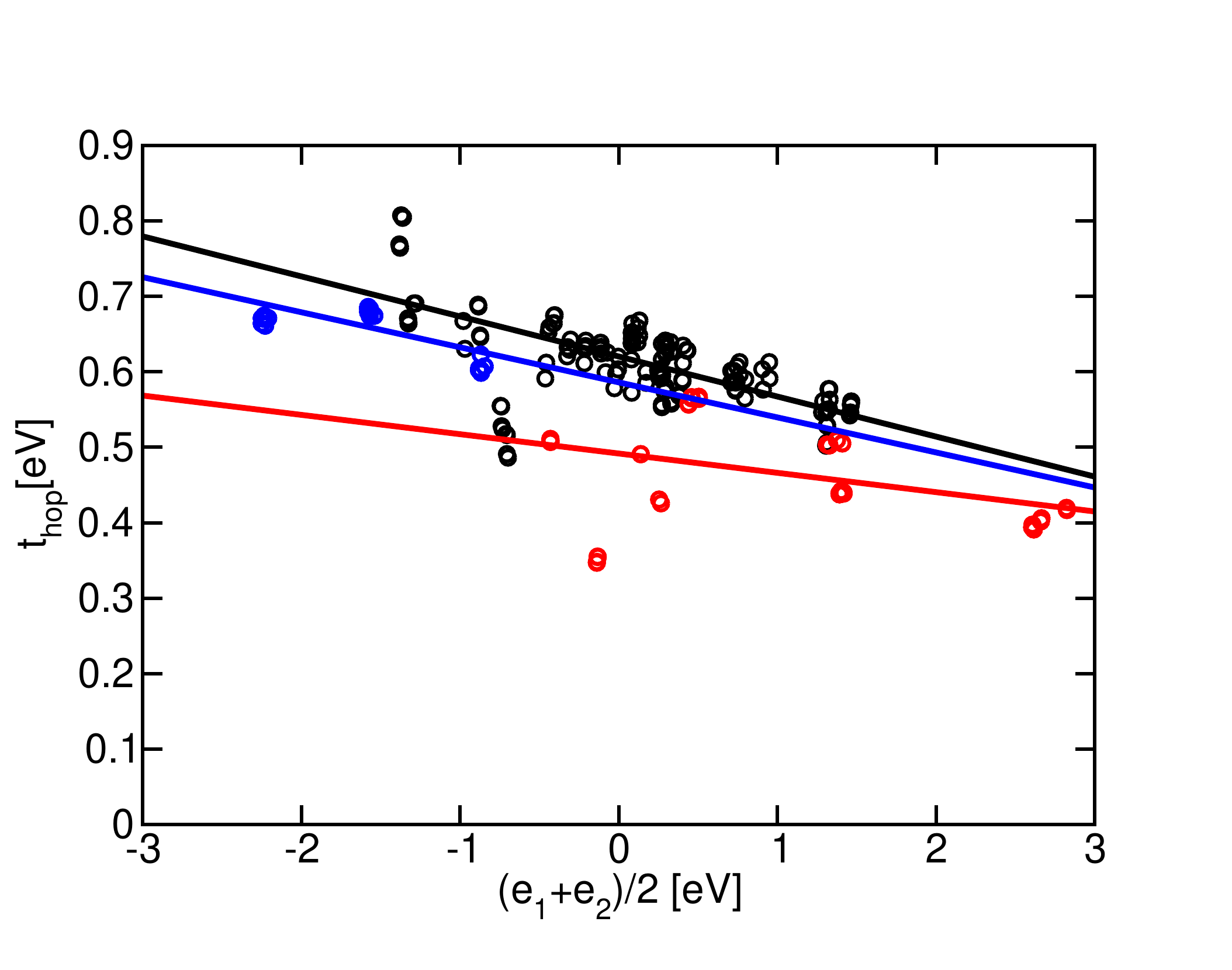}
\end{center}
\caption{\label{fig:thopfit}(Color online) Hopping parameters
  $t_{hop}$ extracted from the generalized DoS for {PrMnO$_3$} (red),
  {Pr$_{1/2}$Ca$_{1/2}$MnO$_3$} (black) and {CaMnO$_3$} (blue) as
  function of the mean orbital energies $\epsilon_1$ and $\epsilon_2$
  in a bond. The straight lines are linear interpolations.}
\end{figure}

The hopping parameters exhibit a considerable spatial variation,
which, however, seems to follow a, yet unknown, systematics. One trend
is a correlation with the level positions, shown in
figure~\ref{fig:thopfit}.  The level positions are the orbital energies
extracted from the projected DoS via
Eq.~\ref{eq:orthohamiltonfromdosparmslevel}. We obtain the fit
\begin{eqnarray}
t_{hop}=0.585\textrm{eV}-0.552
\left(\frac{\epsilon_1+\epsilon_2}{2}-
\left\langle\frac{\epsilon_1+\epsilon_2}{2}\right\rangle\right)
\end{eqnarray}
where the term in angular brackets is the average over all bonds
considered.  The variation of the hopping terms has been
attributed\cite{kovacik16_prb93_75139} to the fact that the hopping is
mediated by the O-p orbitals. Orbitals closer to the oxygen valence
band thus experience a stronger hybridization than more distant
orbitals.

The parameter set, which results from the procedure described above,
is shown in table~\ref{tab:parameters}
\begin{table}[!ht]
\caption{\label{tab:parameters} Parameters for the model described in
  section~\ref{sec:model} extracted from first-principles calculations
  and the values of the underlying physical quantities.}
\begin{center}
\begin{tabular}{lrl|lrl|lrl}
\hline
\hline
  $J_H$          & 0.653&eV 
& $g_{br}$       & 2.988 &eV/\AA               
& $\Delta_H(0)$  & 1.306 &eV \\
  $U$            & 2.514 &eV 
& $k_{br}$       & 10.346 &eV/\AA$^2$   
&$\Delta_H(1)$   & 2.909 &eV\\
  $J_{xc}$       & 0.692 &eV
& $J_{AF}$       & 0.014  &eV               
& $\Delta^\uparrow$& 2.165 &eV\\
  $g_{JT}$           & 2.113  &eV/\AA 
& $t_{hop}$ & 0.585  & eV                
& $\Delta^\downarrow$ & 0.343 &eV\\
  $k_{JT}$         & 5.173  &eV/\AA$^2$ 
& $\bar{d}$& 1.923&\AA 
& $\sqrt{Q_2^2+Q_3^2}$  & 0.409 &\AA\\
\hline
\hline
\end{tabular}
\end{center}
\end{table}

\subsection{Discussion of the parameter set}
\label{sec:parameterdiscuss}
\begin{figure}[!ht]
\begin{center}
\includegraphics[width=\linewidth]{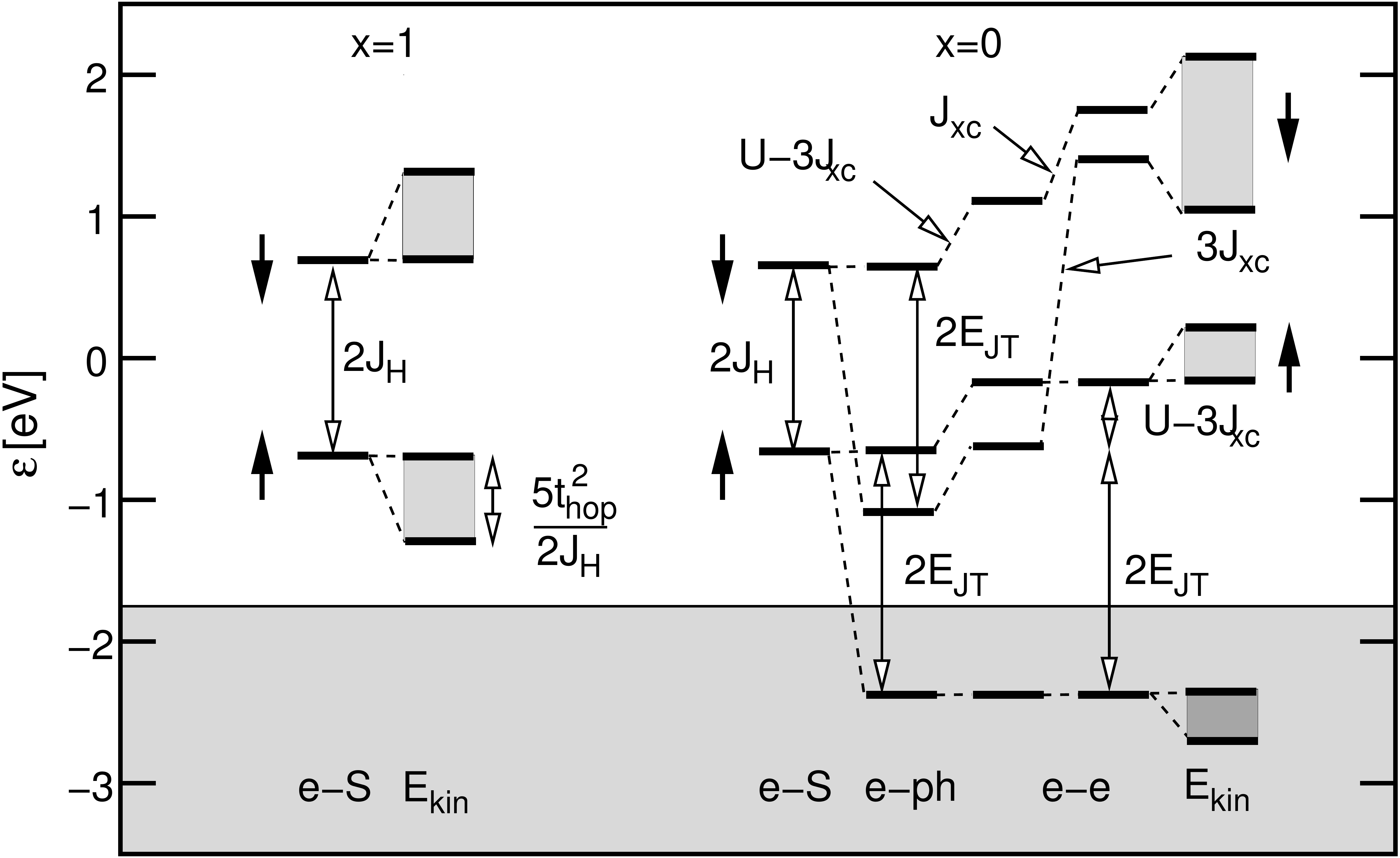}
\end{center}
\caption{\label{fig:levels} Schematic diagram of the energy levels in
  the for {CaMnO$_3$} (left), and {PrMnO$_3$} (right). The onsite
  model with parameters from table~\ref{tab:parameters} is used. The
  effect resulting from the individual terms of the model Hamiltonian
  on the energy levels is shown: Hund's-rule energy $E_{e-S}$ (e-S),
  Jahn-Teller $E_{e-ph}$ (e-ph), and the Coulomb energy $E_U$
  separated into $E_H+E_{SIC}$ and $E_{OR}$ (e-e). The effect of the
  kinetic energy ($E_{kin}$) has been estimated using a second-order
  expression for the intersite hopping.}
\end{figure}

The energy levels resulting from the onsite model for {CaMnO$_3$} (x=1)
and {Pr$_{1/2}$Ca$_{1/2}$MnO$_3$} (x=0) are shown in
figure~\ref{fig:levels}. It demonstrates the relative importance of
the individual terms in the Hamiltonian. The kinetic energy has been
included using perturbation theory for the hopping between
ferromagnetically aligned Mn-sites.

\paragraph{Breathing term:} 
Figure~\ref{fig:levels} makes it evident that, due to the inclusion of
the breathing term, the Jahn-Teller effect lowers one orbital instead
of lowering one orbital and raising the other.

The same ratio $k_{br}/k_{JT}=2$ as the one determined above
  in Eq.~\ref{eq:kbrandkjt} on the theoretical grounds has been
  estimated independently by Hotta\cite{hotta02_arxiv0212_466} on the
  basis of measured phonon frequencies. Our
  first-principles calculations for {PrMnO$_3$} in the A-type
  structure lead to an even larger ratio of breathing amplitude and
  Jahn-Teller amplitude than anticipated by Eq.~\ref{eq:ratioQ123}:
  With the data from table~\ref{tab:strc1}, we obtain values for
  $Q_1=0.402$~{\AA} and $\sqrt{Q_2^2+Q_3^2}=0.406$~{\AA, which are of
    the similar size}.

\paragraph{Coulomb interaction:}
The two Coulomb parameters in our model, $U$ and $J_{xc}$, capture all
Coulomb matrix elements within an $e_g$ orbital shell. 
We considered all four Kanamori
  parameters\cite{kanamori63_progtheorphys30_275}.
 \begin{eqnarray}
  U&=&W_{\alpha,\alpha,\alpha,\alpha} \quad\text{for $\alpha\in\{a,b\}$}
\nonumber\\
  U'&=&W_{\alpha,\beta,\alpha,\beta}=U-2J_{xc}
\quad\text{for $\alpha,\beta\in\{a,b\}$ and $\alpha\neq\beta$}
\nonumber\\
  J&=&W_{\alpha,\beta,\beta,\alpha}=J_{xc}
\quad\text{for   $\alpha,\beta\in\{a,b\}$ and $\alpha\neq\beta$}
\nonumber\\
  J'&=&W_{\alpha,\alpha,\beta,\beta}=J_{xc}
\quad\text{for $\alpha,\beta\in\{a,b\}$ and $\alpha\neq\beta$}
\end{eqnarray}
with 
\begin{eqnarray}
W_{\alpha,\beta,\gamma,\delta}=\int d^3r\int
d^3r'\;
\frac{\chi^*_\alpha(\vec{r})\chi^*_\beta(\vec{r'})
\chi_\gamma(\vec{r})\chi_\delta(\vec{r'})}{4\pi\epsilon_0|\vec{r}-\vec{r'}|}
\end{eqnarray}
For the $e_g$ orbitals, only two parameters $U$ and $J_{xc}$ are
independent.

Our model sheds light on the long-standing discussion about the
  relative importance of Coulomb and Jahn-Teller effect on the orbital
  ordering in manganites. In a seminal paper\cite{kugel73_jetp37_725},
  Kugel and Khomsky showed that orbital ordering can result solely
  from Coulomb interactions, while the Jahn-Teller distortions are
  expected to play a secondary role. Recent calculations have
  established the dominance of the Jahn-Teller
  effect\cite{pavarini08_prl101_266405,flesch12_prb85_35124}. The
  one-site model supports this finding from a somewhat different
  perspective. Let me consider the limit of large
  Hund's coupling, where the spins of the $e_g$ electrons are
  completely aligned with the spin $\vec{S}$ of the $t_{2g}$
  orbitals. This is a reasonable assumption for the manganites under
  study. In this case, the only orbital-dependent terms in the onsite
  model Eq.~\ref{eq:onsitemodelenergy} are the Jahn-Teller and the
  self-interaction energies. Both act in an identical manner so that
  they can be combined into one term
\begin{eqnarray}
E_{e-ph}+E_{ph}+E_{SIC}=-\Bigl(\frac{g_{JT}^2}{k_{JT}}+\frac{1}{2}(U-3J_{xc})\Bigr)
\sum_{\alpha}n_{\uparrow,\alpha}^2
\nonumber\\
\end{eqnarray}
This term favors orbital polarization.  While providing a simplified
view of the problem, this argument shows how the Coulomb interaction
and the Jahn-Teller effect act similarly. With our set of parameters,
the Coulomb interaction accounts for 20~\% of the net effect, while
the dominant effect is due to the Jahn-Teller distortions. The
relative size of the two contributions becomes significant for
relaxation processes, because they are expected to operate on
different time scales.  Compared to electronic relaxation processes,
the response of Jahn-Teller distortions is slow. Thus, we have
provided a very intuitive picture of the Kugel-Khomsky
mechanism\cite{kugel73_jetp37_725}, and a measure for the relative
importance of Kugel-Khomsky and Jahn-Teller mechanisms for orbital
ordering.

The Coulomb interaction does not affect the energetic position of the
occupied orbital in {Pr$_{1/2}$Ca$_{1/2}$MnO$_3$} as seen in
figure~\ref{fig:levels}. This is a consequence of the absence of any
self interaction between electrons in the orbital-ordered
state. 

The total Coulomb repulsion for a density matrix
with exactly one electron vanishes identically.  This case is
particularly relevant for the manganites, because (1) they contain
between zero and one electrons and (2) the correlations tend to favor
integral occupations. The balance between the different terms of the
U-tensor is broken, when only parts of the U-tensor are considered.
Therefore, conclusions drawn on the basis of a partial U-tensor need
to be considered with caution.

Notable is that the Coulomb parameters $U$ and
$J_{xc}$ occur often in a combination $U-3J_{xc}=0.44$~eV, which is
considerably smaller than the intraband Coulomb parameter $U=2.51$~eV.
This reduces the net effect of the Coulomb interaction compared to
approximations that ignore $J_{xc}$. The term proportional to
  $U-3J_{xc}$ dominates, whenever the Mn-sites are fully magnetized,
  which is fulfilled in the manganites to a good approximation.
  $E_{OR}$, in contrast, acts only on the minority-spin states.

The polaron formation energy  including Jahn-Teller and breathing terms is
\begin{eqnarray}
E_{JT}=-\frac{1}{2}\Bigl(
\frac{g_{JT}^2}{k_{JT}}
+\frac{g_{br}^2}{k_{br}}\Bigr)
\label{eq:defpolaronformationenergy}
\end{eqnarray}
Usually, the breathing term in polaron formation energy is ignored,
which makes the comparison of our values with those in the literature
ambiguous. 

In the absence of the breathing term, there is a simple relation
$\Delta_{JT}=2E_{JT}$ between the polaron formation energy $E_{JT}$
and the Jahn-Teller splitting $\Delta_{JT}$ of the energy levels.  If
the breathing term is considered, the relation between the polaron
formation energy and the Jahn-Teller splitting of the electron levels
 is different.

With the definition Eq.~\ref{eq:defpolaronformationenergy}
of the polaron formation energy given above, we obtain
$\Delta_{JT}=E_{JT}=0.86~eV$. 

The dimension-less electron-phonon coupling constant is
$\lambda=\sqrt{2E_{JT}/t_{hop}}$. We obtain $\lambda=1.72$ with and
$\lambda=1.21$ without breathing contribution.
Millis\cite{millis96_prb54_5405} concluded that $\lambda$ lies between
1.3 and 1.5.

The hopping is expected to depend on the bond angle of the bridging
oxygen ion via
$t_{hop}(\theta)=t_{hop}(180^\circ)\cos(\theta-180^\circ)$.  The
dependence of the hopping parameter on the bond angle, explains its
correlation with the tolerance factor and the resulting octahedral
tilt. Due to the similar ionic radius of Ca and Pr, the bond angle
variation in the materials considered here is relatively small.  Our
hopping parameter $t_{hop}=0.585$ extrapolated to $180^\circ$ yields
$t_{hop}(180^\circ)=0.65$~eV.

The smaller splitting of the minority-spin energy levels as compared
to the majority-spin levels has been attributed to reduced
bond-strength with the oxygen orbitals.\cite{ederer07_prb76_155105}
This general trend is confirmed in
figure~\ref{fig:thopfit}. Nevertheless, our model uses a uniform
hopping parameter.

For the uniform hopping parameter given above, we find that the model
calculations favor the ferromagnetic structure over the CE-type
antiferromagnetic structure in {Pr$_{1/2}$Ca$_{1/2}$MnO$_3$} by
0.037~eV per Mn site, when using $J_{AF}=3.26$~meV. This is
inconsistent with the experimental finding.

This difficulty can be healed by increasing the antiferromagnetic
coupling beyond $J_{AF}=14$~meV. Raising the antiferromagnetic
coupling is a common cure for this problem, which does not affect the band
structure. 

Besides the dependence on $J_{AF}$, also a reduction of the hopping
parameter in the ferromagnetic structure by 0.08~eV is sufficient to
interchange the stability. This is within the range of the scattering
of the hopping parameters extracted from the ab-initio
calculations. Indeed we found the largest hopping parameters within
the zig-zag chain of the CE-type structure, while the ferromagnetic
calculations had unusually small hopping parameters. A more accurate
description of the hopping parameters requires more work.

Because Jahn-Teller distortion and hopping are related to the
hybridization of Mn-d states with the p-levels on the bridging oxygen,
it has been suggested\cite{kovacik16_prb93_75139} to reduce the
Jahn-Teller splitting for the minority-spin direction. Our model
attributes this reduction to the onsite Coulomb interaction.

The same ratio $k_{br}/k_{JT}=2$ as the one determined above has been
estimated independently by Hotta\cite{hotta02_arxiv0212_466} on the
basis of the ratio of phonon frequencies. With $\bar{d}=1.923$~{\AA}
extracted from a calculation for {CaMnO$_3$}, we find for {PrMnO$_3$} in the
A-type structure $Q_1=0.402$ and $\sqrt{Q_{2}^2+Q_3^2}=0.409$~\AA,
which indicates that the breathing distortion may be even larger than
the ratio of Eq.~\ref{eq:ratioQ123} in the current parameterization.

\section{Electronic structure of {Pr$_{1-x}$Ca$_{x}$MnO$_3$}}
The gross features of the electronic structure have been discussed
earlier. Here we will focus on finer details and relate them to
absorption spectra.

\subsection{{CaMnO$_3$}}
\label{sec:cmo}

The calculated direct band gap for {CaMnO$_3$} of 1.58~eV agrees well with
the onset of our measured optical conductivity at 1.45~eV and other
measurements at 1.55~eV.\cite{loshkareva04_prb70_224406} The
calculated fundamental band gap is with 1.47~eV slightly smaller than
the direct gap. In {CaMnO$_3$}, we expect at room temperature a thermal
average of different magnetic orders due to their near
degeneracy. Such an average, however, would not affect the band gap
strongly, because our calculations predict similar direct gaps in the
range from 1.47~eV to 1.58~eV for the G-, C-, and A-type
antiferromagnetic orders and fundamental band gaps in the range from
1.41~eV to 1.47~eV. The fundamental gaps become relevant if the
magnetic disorder is very short ranged, so that the k-selection rule is
no more valid.

\begin{figure}[!htb]
\begin{center}
\includegraphics[width=\linewidth,clip]{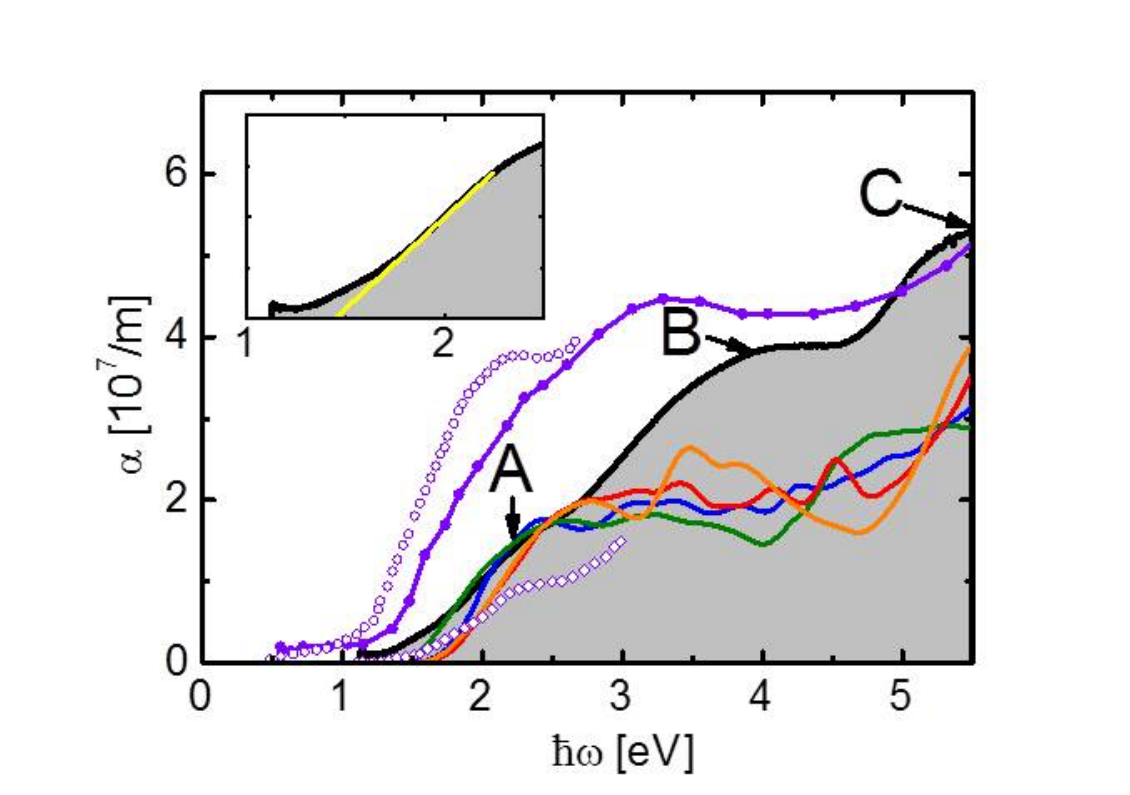}
\end{center}
\caption{\label{fig:optabsorbcmo} Measured and calculated absorption
  coefficients of {CaMnO$_3$}. Our room-temperature measured data for
  $x=0.95$ are shown as grey-shaded area. Shown are also the results
  by Jung et al.\cite{jung97_prb55_15489} (solid symbols) and Asanuma
  et al.\cite{asanuma09_prb80_235113} (open symbols). The calculated
  spectra are not to scale. They are shown for G-type (orange) C-type
  (red), A-type blue and B-type (green) magnetic order. The inset
  illustrates the determination of the optical gap of 1.45~eV via
  linear extrapolation of the low energy tail of the spectrum}
\end{figure}

Measured absorption coefficients are compared with the calculated
contributions of selected transitions in
figure~\ref{fig:optabsorbcmo}.  This comparison provides a detailed
interpretation of the observed features.

Our experimental spectrum in figure~\ref{fig:optabsorbcmo} exhibits
three shoulders, namely shoulder (A) at 2.2~eV, shoulder (B) at 4~eV
and shoulder (C) at 5.5~eV. The spectrum by Jung et
al.\cite{jung97_prb55_15489} show the same qualitative behavior.  The
analysis of Asanuma et al.\cite{asanuma09_prb80_235113} indicates that
changes in the oxidation state have a considerable impact on the
intensity of the spectra and the energetic position of the main
features, which we will discuss in the next section.

The calculated spectra exhibit, for all magnetic orders, a sharp rise
from the conduction band edge to a shoulder at about 2.5-3~eV, which
we identify with shoulder (A) of the measured spectrum. The shoulder
is followed by a quasi plateau that rises sharply again beyond
5~eV. We attribute this rise with shoulder (C). Shoulder (B) in the
experimental spectrum can be attributed to the corresponding peak in
the calculated spectrum at 3.5~eV for the G-type magnetic order.

\begin{figure}[!htb]
\begin{center}
\includegraphics[width=0.5\linewidth]{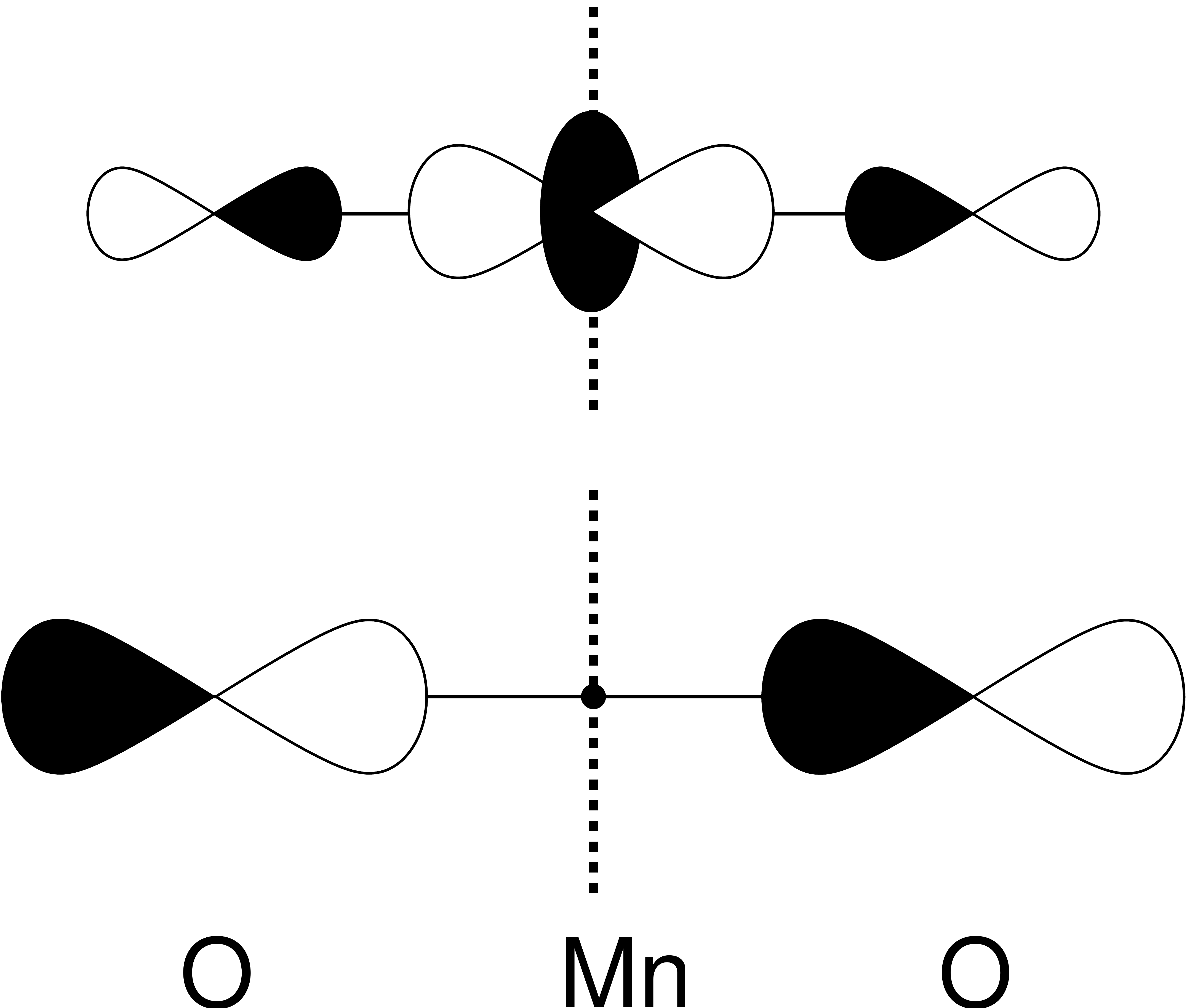}
\end{center}
\caption{\label{fig:cmoexit}Diagram of O-p state (bottom) in the
  valence band and the Mn-$e_g$ state (top) in the conduction band
  responsible for the charge-transfer transition in {CaMnO$_3$}.  The
  dashed line indicates the reflection plane defining the parity for
  the dipole selection rule. Each MnO$_6$ octahedron has three such
  orbitals, each active for another polarization direction.}
\end{figure}

The optical absorption in {CaMnO$_3$} is due to charge-transfer transitions
from O-p orbitals in the oxygen valence band into Mn-$e_g$ states.
The spectra are calculated as described in appendix~\ref{app:optabs}.
The participating orbitals responsible for the absorption are sketched
in figure~\ref{fig:cmoexit}.  

The feature (B) at 3.5~eV is present only in the calculated spectra
for {CaMnO$_3$} with the G-type magnetic order.  Like shoulder (A), the
absorption is due to a transition into the lower band of the Mn-d
states dominated by the majority-spin $e_g$-states and minority-spin
$t_{2g}$ states. The subsequent depletion in the absorption at 4.7~eV
has two origins: firstly, it is due to the band gap in the final state
DoS at 4~eV in figure~\ref{fig:cmo_dos}, which separates minority-spin
from majority-spin $e_g$ states; secondly, it reflects the quasi gap
in the initial state DoS at -5~eV in figure~\ref{fig:cmo_dos}.  The
absorption spectrum can be considered as two replicas of the broadened
final density of states, scanned once by the upper part of the oxygen
valence band and then, separated by 2-3~eV, a replica scanned by the
lower part of the oxygen valence band. Consequently, the rise in the
absorption coefficient above 5~eV is due to transitions into the
minority-spin and the majority-spin $e_g$ states. While this
assignment is consistent with that of Jung et
al.\cite{jung97_prb55_15489}, it also accounts for the structure in
the initial-state DoS.

\begin{figure}[!htb]
\begin{center}
\includegraphics[width=\linewidth,clip]{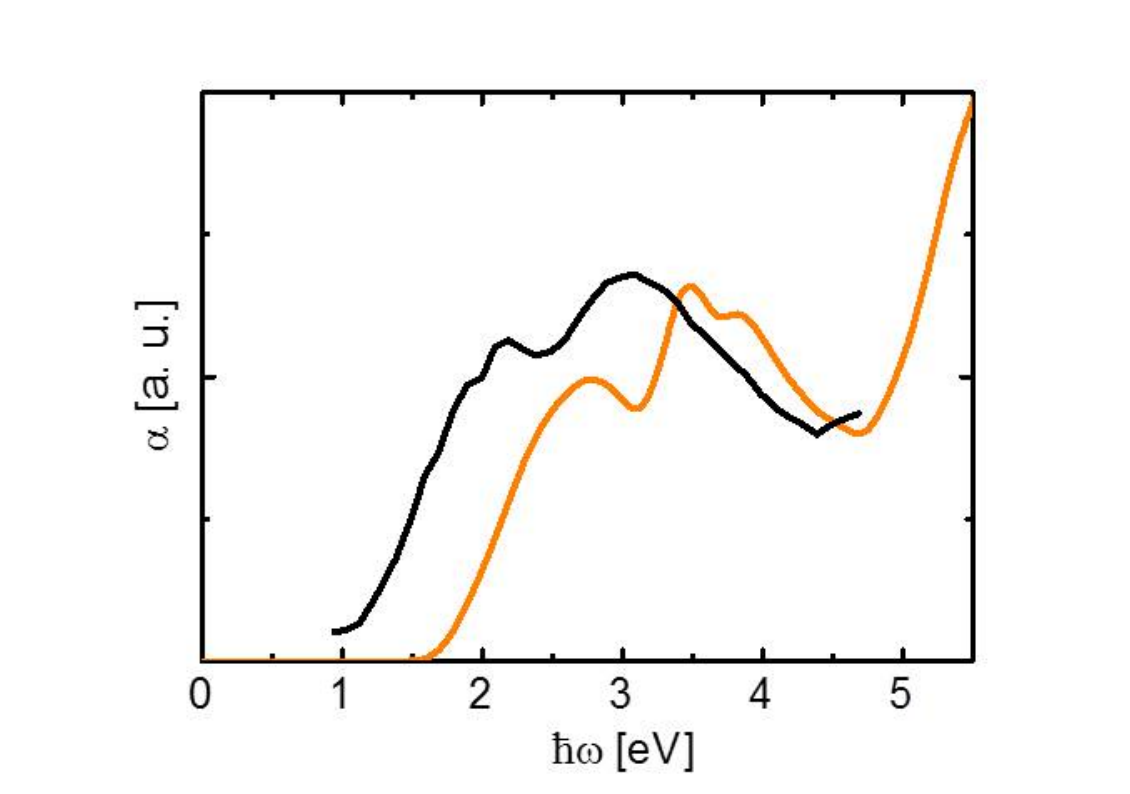}
\end{center}
\caption{\label{fig:optabsorbcmolosh}(Color online) Measured
  room-temperature absorption spectra by Loshkareva et
  al.\cite{loshkareva04_prb70_224406} for {CaMnO$_3$} (black) and
  calculated spectra for G-type {CaMnO$_3$}. The calculated data
  (orange, grey) have been scaled independently to allow comparison of
  the spectral shapes. }
\end{figure}

Most conspicuous is that the peak (B) and the minimum in the
calculated absorption for G-type {CaMnO$_3$} at 4.7~eV is represented only
as a plateau in the experiment.  While it is absent in our absorption
spectra and very weak in the photoconductivity by Jung et
al.\cite{jung97_prb55_15489}, it is clearly seen in the
photoconductivity spectra of Loshkareva et
al.\cite{loshkareva04_prb70_224406} shown in
figure~\ref{fig:optabsorbcmolosh}.

A possible explanation for the different shapes of the experimental
spectra is that not only the G-type antiferromagnetic order, but also
other antiferromagnetic orders such as C- and A-type magnetic order
contribute considerably to the thermal ensemble. Due to the increased
band width of the Mn-$e_g$ states in the C and A-type order, the
depletion in the calculated absorption spectrum at 4.7~eV is completely
washed out and a shoulder occurs instead.  Given the small energy
difference between magnetic orders of {CaMnO$_3$}, also other effects such
as strain and preparation-induced point defects in the thin films can
influence the distribution of magnetic orders and thus cause different
absorption spectra.  An improved understanding of the spectra may thus
provide better control of the samples.

In our samples 5~\% of Ca ions on the A-site are substituted by Pr
ions. Thus they have a certain number of electron polarons. Electrons
in {CaMnO$_3$} favor an environment with ferromagnetically aligned Mn ions,
because this lowers the kinetic energy of electron polarons. Thus,
a small doping of {CaMnO$_3$} shifts the stability away from the G-type
towards the C- and A-type antiferromagnetic orders. Therefore, the
remaining doping in our probes with $x=0.95$ enhances the other
antiferromagnetic orders and thus washes out the trough in the
spectrum at 4.7~eV.

The agreement of the shape of our calculated optical absorption
spectra of G-type {CaMnO$_3$} with the optical conductivity measured by
Loshkareva et al.\cite{loshkareva04_prb70_224406}, shown in
figure~\ref{fig:optabsorbcmolosh} is striking. Given that these
structures are not present in calculated spectra of the other magnetic
orders investigated, it may indicate a stronger contribution of G-type
antiferromagnetic ordering to the thermal average in these
experimental probes.

\subsection{{PrMnO$_3$}}
\label{sec:pmo}
\begin{figure}[!htb]
\begin{center}
\includegraphics[width=\linewidth,clip]{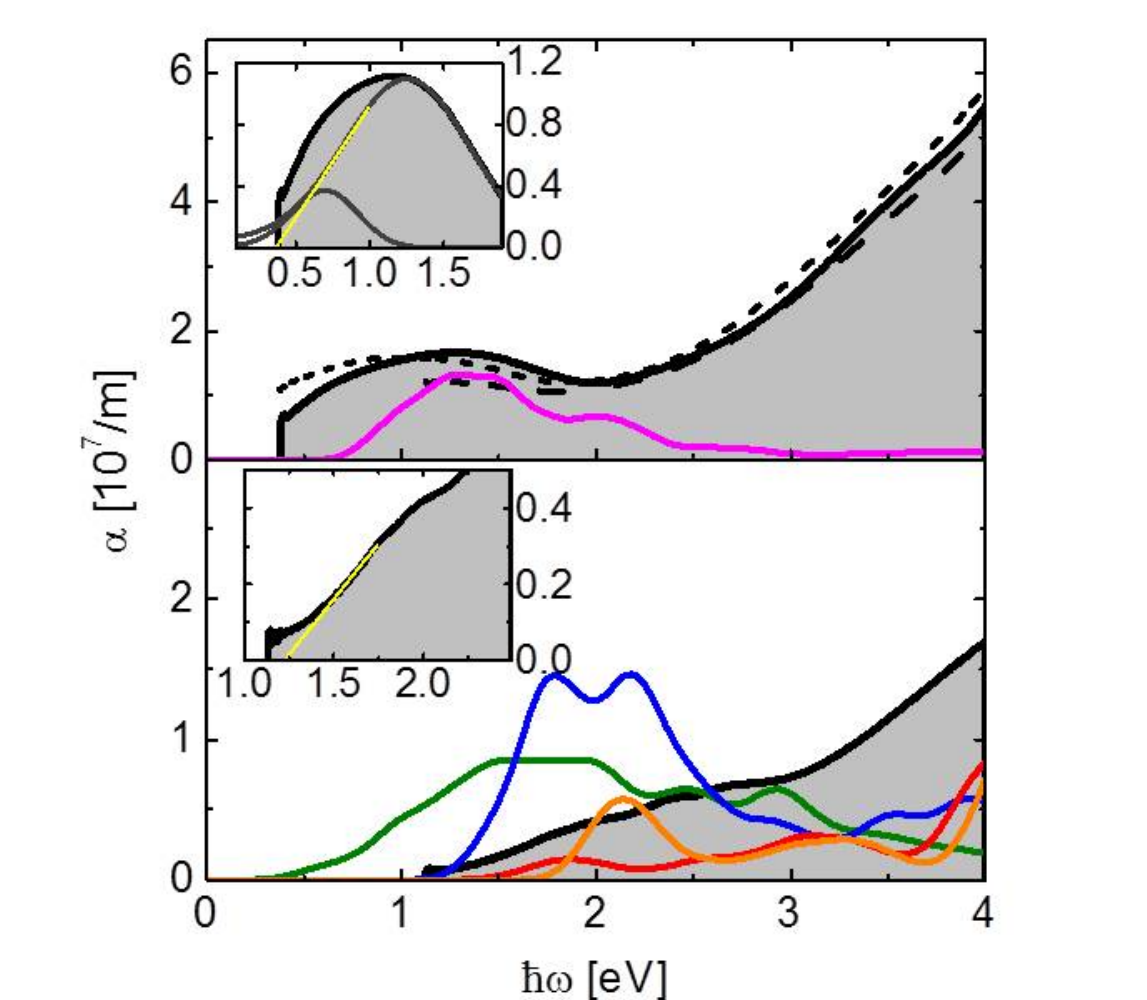}
\end{center}
\caption{\label{fig:optabsorb} Absorption coefficients of
  {Pr$_{1/2}$Ca$_{1/2}$MnO$_3$} (top) and {PrMnO$_3$} (bottom).
  Measured data are indicated by the black line and the shaded area.
  Calculated spectra for d to d transitions have been performed for
  the G-type (orange), C-type (red), A-type (blue), B-type (green) and
  CE-type (magenta) magnetic orders.  Measurement for
  {Pr$_{1/2}$Ca$_{1/2}$MnO$_3$} have been performed at 300~K with
  (short-dashed line) and without (long-dashed line) reflection
  correction and at 80~K without (full line and shaded area)
  reflection correction. Spectra for {PrMnO$_3$} have been done at
  300~K with reflection correction. Insets show the extrapolation used
  for the optical gap determination. For details, see text.}
\end{figure}

The optical gap of 1.25~eV for {PrMnO$_3$} extracted from our measured
optical conductivity in figure~\ref{fig:optabsorb} is well reproduced
by our calculated direct gap of 1.30~eV for the A-type magnetic
order. The calculated fundamental band gap is smaller, namely 1.05~eV.

The center of the calculated absorption band at 2~eV for the A-type
order shown in figure~\ref{fig:optabsorb} is lower than that of the
measured absorption band of 2.4~eV.\cite{mildner15_prb92_35145}
Furthermore, our calculations for the A-type magnetic order
overestimate the measured optical absorption. For the other
antiferromagnetic orders, the absorption is considerably smaller. In
particular, the absorption of the G-type magnetic order agrees better
in peak position and magnitude.  We therefore conclude that the A-type
magnetic order is not dominant under the experimental
conditions. Conceivable is that a fluctuating antiferromagnetic order
is present.

\begin{figure}[!htb]
\begin{center}
\includegraphics[width=0.8\linewidth]{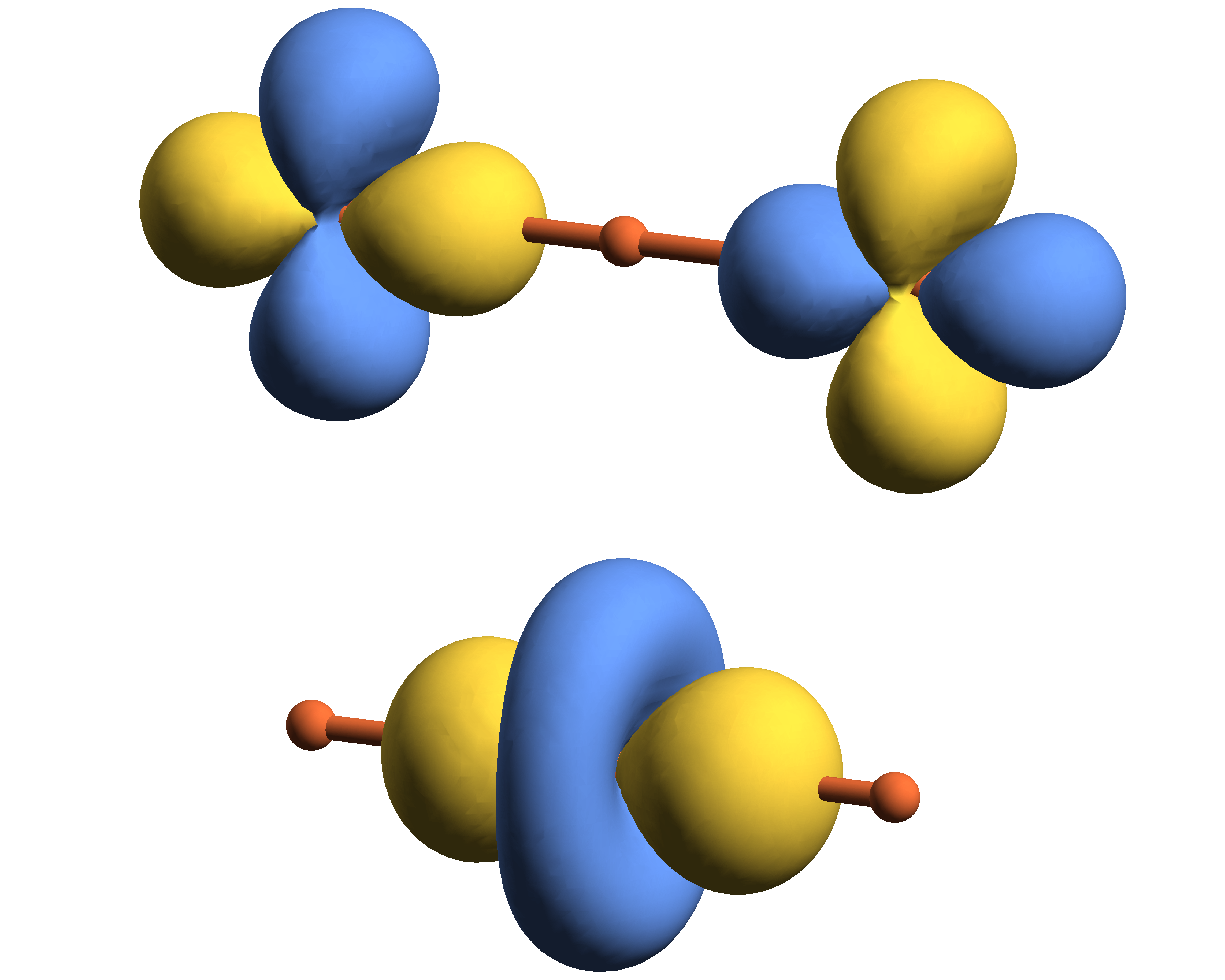}
\end{center}
\caption{\label{fig:pmoexcit}(Color online) Orbitals involved in the
  optical transition in {PrMnO$_3$} from the filled (lower graph) to
  the empty Jahn-Teller split band (upper graph). Only the Mn-centered
  $e_g$ orbitals are shown. Orbitals are shown with the calculated
  orbital-mixing angle $\gamma=55^\circ$ of the occupied orbital. The
  octahedral tilt is ignored. The octahedron on the central site is
  expanded along the axis, while the octahedra on the terminal sites
  are expanded in the perpendicular direction. }
\end{figure}

The optical-absorption line centered at
2.4~eV\cite{mildner15_prb92_35145} has been attributed to the
transition from the lower to the upper Jahn-Teller
band,\cite{kim06_prl96_247205} as sketched in
figure~\ref{fig:pmoexcit}.  While the onsite d-to-d transition is
dipole forbidden, a transition between two $e_g$ orbitals centered at
different sites is dipole allowed.

Let us investigate the electronic structure in greater detail: The
lower, filled states $|w_{l}\rangle$ of {PrMnO$_3$} can be expressed in
terms of cubic harmonics as
\begin{eqnarray}
|w_{l}\rangle&=&
-|d_{3z^2-r^2}\rangle \cos(\gamma)
+|d_{x^2-y^2}\rangle \sin(\gamma)
\label{eq:pmoocc}
\end{eqnarray}
where our calculations yield an mixing angle $\gamma=55^\circ$ in
A-type {PrMnO$_3$}. 

The corresponding upper, unoccupied orbitals
$|w_{u}\rangle$ are of the form
\begin{eqnarray}
|w_{u}\rangle&=&
-|d_{3z^2-r^2}\rangle \sin(\gamma)
-|d_{x^2-y^2}\rangle \cos(\gamma)
\label{eq:pmoempty}
\end{eqnarray}

The orbitals on nearest-neighbor Mn sites are rotated by 90$^{\circ}$
about the local z-axis, which is accomplished by changing the sign of
$|d_{x^2-y^2}\rangle$ in Eq.~\ref{eq:pmoocc} and
\ref{eq:pmoempty}. The orbital-mixing angle angle $\gamma=55^\circ$ is
close to $45^\circ$ for which filled and empty orbitals have the same
shape except for translations and $90^\circ$ rotations about the
z-axis. The orbital-mixing angle $\gamma$ is the same for all
antiferromagnetic orders, A-, C-, and G-type. For the ferromagnetic
B-type order it is, with $60^\circ$, slightly larger.

In figure~\ref{fig:pmo_dos}, the DoS of {PrMnO$_3$} is shown alongside
with the DoS projected separately onto each of the two
Jahn-Teller-split orbitals. This figure shows how well the filled and
empty Jahn-Teller bands can be represented by a single orbital each.

The dominant excitation in the near-infrared (NIR) region at about
2~eV is from an occupied orbital, Eq.~\ref{eq:pmoocc}, to an
antisymmetric combination of the empty orbitals,
Eq.~\ref{eq:pmoempty}, on the neighboring Mn-sites. The two orbitals
are shown in figure \ref{fig:pmoexcit}.

The matrix elements calculated as described in
appendix~\ref{app:optabs} are largest along the extended axis of the
octahedron. The matrix element along the $c$-direction is about half
this value. The matrix element for a polarization in the $ab$-plane
orthogonal to the extended axis of the octahedron practically
vanishes.

As discussed by Kim et al.,\cite{kim06_prl96_247205} the transition is
only considerable between ferromagnetically coupled Mn sites, while
the transition between antiferromagnetically coupled Mn sites,
i.e. along the $c$-axis, requires to overcome Hund's-rule splitting,
which places the absorption energy into the high-energy region of
3-6~eV. Its contribution to the low-energy features has been
considered negligible.

Our quantitative calculations show that the picture is more subtle.
The measured absorption near 2~eV is smaller than expected for
excitations between neighbors of equal spin orientation: the
absorption for A-type and B-type magnetic order at 2~eV is much larger
than the experimental spectra, while the results for the G-type order
are in reasonable good agreement, in terms of both, energetic position
and weight. While the dominant absorption of the G-type structure is
related to the Hund's-rule splitting and occurs above 4~eV, tunneling
of the upper Jahn-teller band into neighboring sites contributes to
the DoS of the minority-spin direction.  It is these wave-function
tails that dominate the optical absorption in the NIR region. This
interpretation gains further support from the comparison of absorption
for {PrMnO$_3$} and {Pr$_{1/2}$Ca$_{1/2}$MnO$_3$}. The optical
absorption in {Pr$_{1/2}$Ca$_{1/2}$MnO$_3$} is considerably larger,
despite the smaller number of optically active orbitals.

\begin{table*}[t]
 \caption{\label{tab:strc1}\label{tab:energies}Mean Mn-O bond-length
   $d_{Mn-O}$, Jahn-Teller distortion $Q=\sqrt{Q_2^2+Q_3^2}$, Mn-O-Mn
   bond angle for oxygen bridges in the $ab$-plane $\phi_{O(2)}$ and
   along the $c$-axis $\phi_{O(1)}$, magnetic moment $\mu$ and number
   of electrons in the Mn-d-shell.  The E-type magnetic order of
   {Pr$_{1/2}$Ca$_{1/2}$MnO$_3$} uses the experimental lattice
   constants for $x=0.4$.\cite{daoudaladine02_prl89_97205} Calculated
   relative energies and fundamental band gaps in
   {Pr$_{1-x}$Ca$_{x}$MnO$_3$} for different magnetic orders and
   doping. For each doping the most stable configuration has been
   chosen as the energy zero.}
\begin{center}
\begin{tabular}{c >{\centering}m{1.1cm} cccccc >{\centering}m{2cm} >{\centering\arraybackslash}m{1cm}}
\hline
\hline
Compound  & Magnetic order & $d_{Mn,O}$ [\AA] & $Q$ [\AA] &  $\phi_{O(1)}$ & $\phi_{O(2)}$ & $\mu[\mu_B]$ & $N_d$ & Energy/$ABO_3$ (meV) & Band gap (eV)\\
         \hline
 {CaMnO$_3$}  & G  &  1.927 & 0.016  & 151.0$^\circ$ & 150.3$^\circ$ & 2.837 & 4.791 & 0   & 1.47 \\
         & C  &  1.927 & 0.012  & 150.4$^\circ$ & 150.4$^\circ$ & 2.853 & 4.793 & -1  & 1.42 \\
         & A  &  1.928 & 0.018  & 151.5$^\circ$ &149.6$^\circ$  & 2.884 & 4.793 & 7   & 1.41 \\
         & B  &  1.929 & 0.014  & 150.4$^\circ$ & 149.7$^\circ$ & 2.922 & 4.794 & 20  & 1.33 \\
         \hline
 {Pr$_{1/2}$Ca$_{1/2}$MnO$_3$} & G  & 2.003/1.933 & 0.056/0.100 & 148.6-156.0$^\circ$ & 147.5-153.2$^\circ$ &  3.686/2.717 &  4.887/4.811 & 104 & 0 \\
         & C  & 1.940/1.997 & 0.098/0.122 & 150.5-154.1$^\circ$ & 148.9-151.5$^\circ$ & 2.885/3.636 & 4.818/4.884  & 96  & 0 \\
         & A  & 1.954/1.979 & 0.056/0.100 & 151.5-154.4$^\circ$ & 149.8-152.5$^\circ$ & 3.299/3.487 & 4.835/4.869 & 54  & 0 \\
         & B  &1.966/1.966  & 0.020/0.016 & 150.6-153.7$^\circ$ & 150.5-153.0$^\circ$ & 3.474/3.481 & 4.856/4.857 & 29  & 0 \\
         & CE & 1.991/1.943 & 0.259/0.053 & 156.0-147.8$^\circ$ & 146.6-157.1$^\circ$ & 3.526/3.125 & 4.880/4.823 & 0   & 0.57 \\
         & E  & 2.014/1.951 & 0.244/0.044 & 149.5-153.8$^\circ$ & 146.5-154.9$^\circ$ & 3.634/2.975 & 4.880/4.813  & - & - \\
\hline

 {PrMnO$_3$}  & G  &  2.044 & 0.470 & 148.3$^\circ$ & 146.3$^\circ$ & 3.662 & 3.662 & 22  & 1.31  \\
         & C  &  2.045 & 0.460 & 147.7$^\circ$ & 146.5$^\circ$ & 3.702 & 4.903 & 29  & 0.90 \\
         & A  &  2.039 & 0.406 & 148.2$^\circ$ & 147.8$^\circ$ & 3.772 & 4.903 & 0   & 1.05 \\
         & B  &  2.039 & 0.392 & 147.7$^\circ$ & 148.0$^\circ$ & 3.819 & 4.902 & 10  & 1.33 \\
 \hline
 \hline
\end{tabular}
\end{center}
\end{table*}

\begin{figure}[!htb]
\begin{center}
\includegraphics[width=\linewidth,clip]{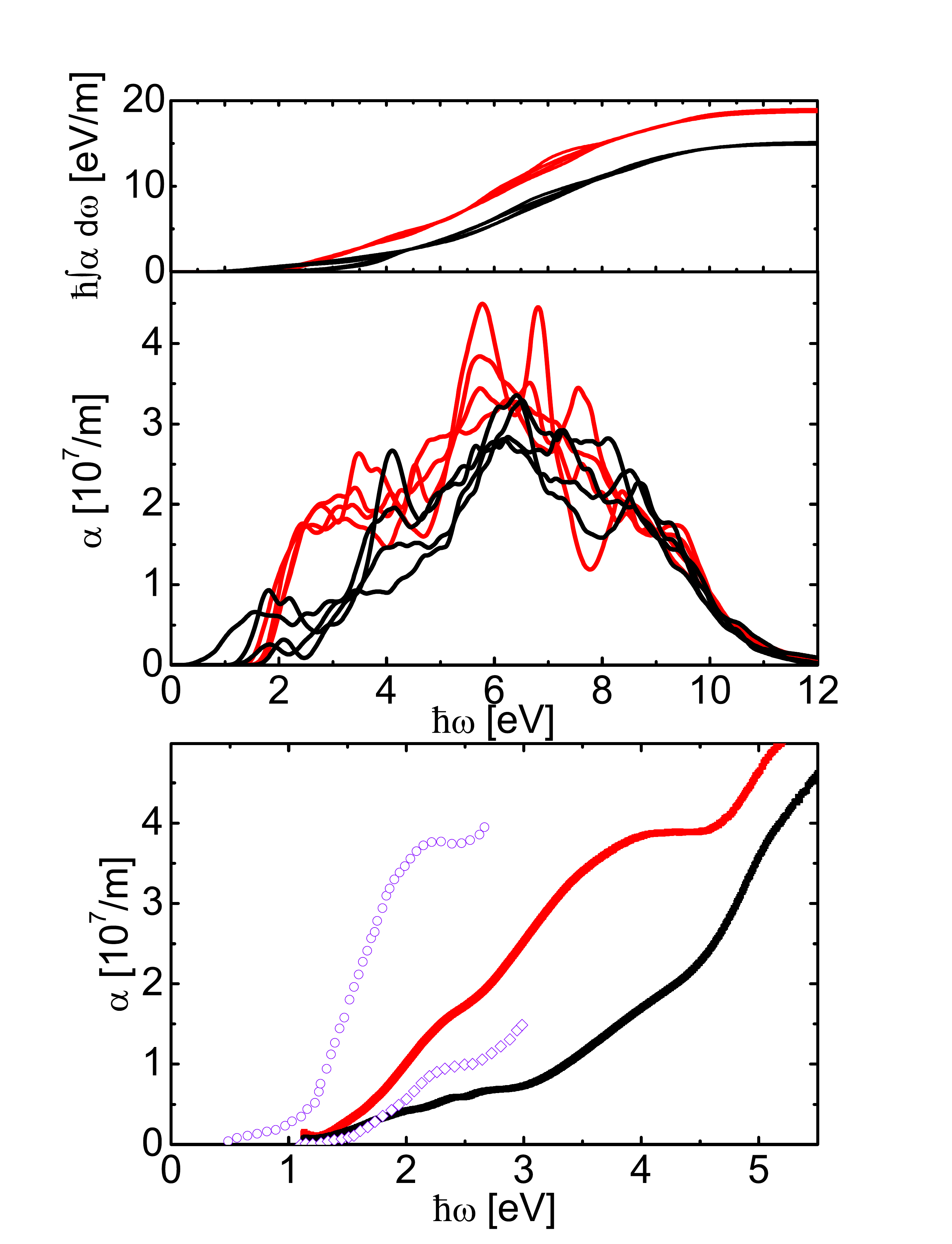}
\end{center}
\caption{\label{fig:optctedge}(Color online) Calculated absorption
  spectra (center) due to charge transfer excitations for {CaMnO$_3$}
  (red, grey) and {PrMnO$_3$} (black). Results are shown for different
  magnetic orders. The apparent right-shift of the absorption edge
  from 2~eV to 4~eV is attributed the occupation of the Mn-$e_g$
  states, which disappear from the spectrum in {PrMnO$_3$}, while they
  are empty and thus visible in the spectrum of {CaMnO$_3$}. The top
  figure shows the integrated absorption spectra, which demonstrates
  the loss of absorption intensity due to occupation of Mn-$e_g$
  states.  The bottom figure shows the experimental results for
  {CaMnO$_3$} (red, grey) and {PrMnO$_3$} (black), which demonstrate
  the lowering of the absorption intensity due to doping. The data by
  Asanuma\cite{asanuma09_prb80_235113} for oxidized (circles) and
  reduced (diamonds) {CaMnO$_3$} exhibit a similar lowering upon
  reduction, which partially occupies the Mn-$e_g$ shell as in the
  {Pr$_{1-x}$Ca$_{x}$MnO$_3$} series.}
\end{figure}

The absorption edge of the charge-transfer transitions in {PrMnO$_3$}
exhibit a considerable shift towards higher energies as compared
  to {CaMnO$_3$}. Only part of this effect can be
attributed to a Coulomb shift.  Comparing the calculated absorption
spectra of {PrMnO$_3$} and {CaMnO$_3$}, there is a marked loss of
absorption intensity between 2~eV and 4~eV, as shown in
figure~\ref{fig:optctedge}. This finding does not depend on the
magnetic order considered.  The loss of intensity can be attributed to
reduction of the number of empty Mn-$e_g$ states by one quarter from
{CaMnO$_3$} to {PrMnO$_3$}. As states become occupied due to doping,
they become unavailable as final states for an optical excitation. The
states in the lower part of the Mn-$e_g$ spectrum are the ones that
are filled, which explains, why this part disappears from the
spectrum. Thus, rather than a shift of the d-states, what is observed
is a depletion of the lower part of the spectrum due to
charge-transfer transitions.

This analysis sheds light onto the large variation of the magnitude of
the measured spectra of {CaMnO$_3$}. Experimentally obtained spectra differ
by a factor 2-4.  In a careful analysis Asanuma et
al.\cite{asanuma09_prb80_235113} has shown that the intensity of the
absorption spectra of {CaMnO$_3$} differ considerably for oxidized and
reduced samples. Reduced samples contain oxygen vacancies, which
introduce electrons into the Mn-d shell. As in {PrMnO$_3$}, these electrons
occupy the lower Mn-d states and, thus, make them unavailable as final
states for charge-transfer transitions. Thus, the apparent variation of
the absorption intensity can be traced back to a microscopic picture.

\subsection{Half-doped manganite {Pr$_{1/2}$Ca$_{1/2}$MnO$_3$}}
In the half-doped Pr$_{1/2}$Ca$_{1/2}$MnO$_3$, the Jahn-Teller
polaron, which dominates the electronic structure of {PrMnO$_3$}, is
replaced by a Zener polaron, which is characterized by an electron
shared by two ferromagnetically coupled Mn neighbors.

Characteristic for a Zener polaron are two neighboring Mn sites, both
having a Jahn-Teller expansion along the axis of the pair.  This is
opposite to the orbital ordering in {PrMnO$_3$}, which exhibits an
alternating series of prolate and oblate octahedral distortions in the
$ab$-plane direction.

While the band gap in {PrMnO$_3$} is mostly due to Jahn-Teller
splitting, in {Pr$_{1/2}$Ca$_{1/2}$MnO$_3$} the lower Jahn-Teller band
is itself split into two, of which only one is occupied.

These two bands can be attributed to bonding
  and antibonding states with respect to the oxygen bridge.  More
precisely, a three-center bond is formed between two neighboring
Mn-sites and a bridging oxygen ion. A three-center bond consists of a
lower, bonding orbital, non-bonding orbital in the middle and an upper
antibonding orbital. The bonding state is energetically located at the
bottom of the oxygen valence band. The non-bonding state is the filled
state of the Zener polaron and the antibonding state of the
three-center bond remains unoccupied in the half-doped
material. 

\begin{figure}[!ht]
\begin{center}
\includegraphics[height=0.4\linewidth]{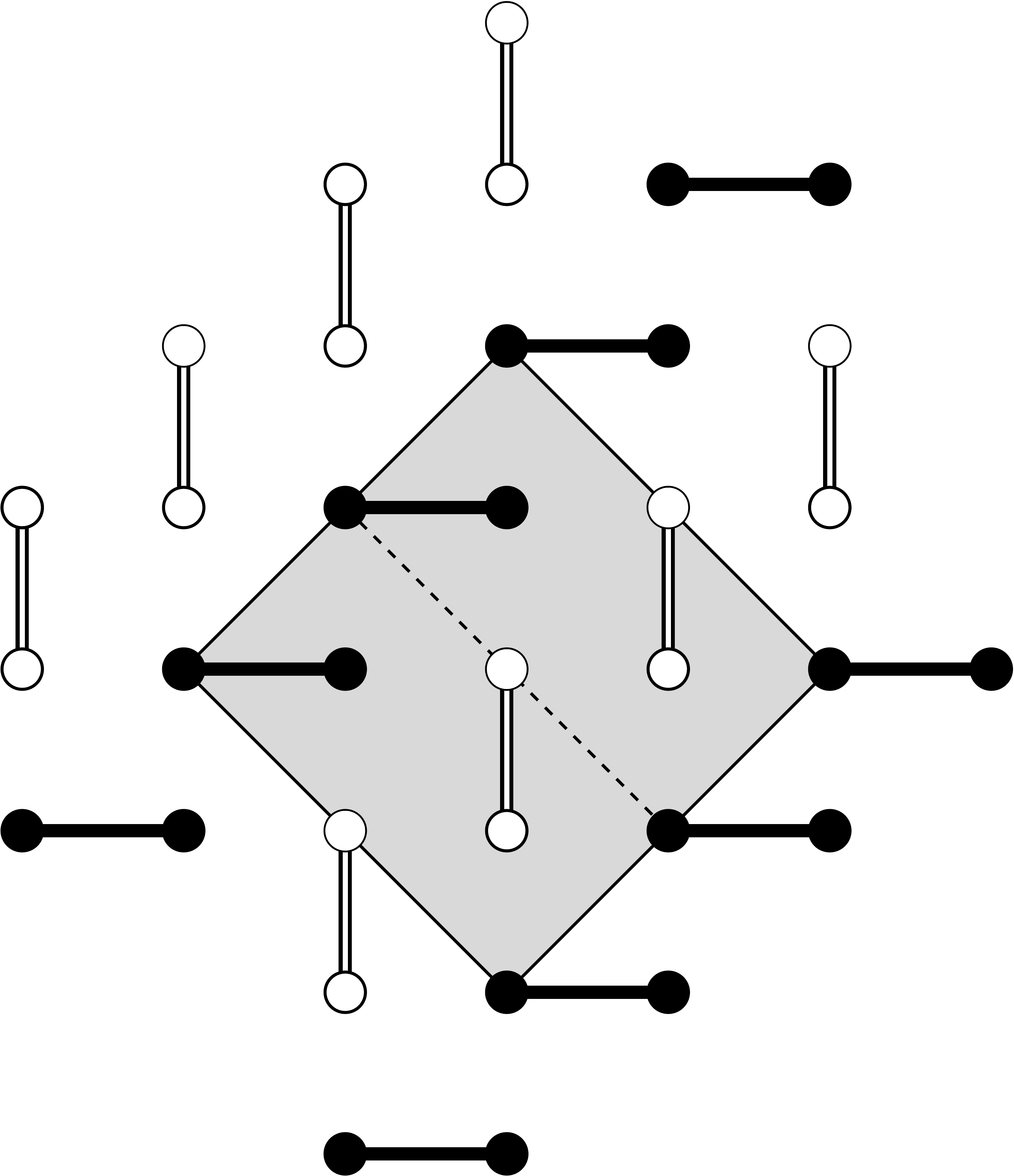}
\includegraphics[height=0.4\linewidth]{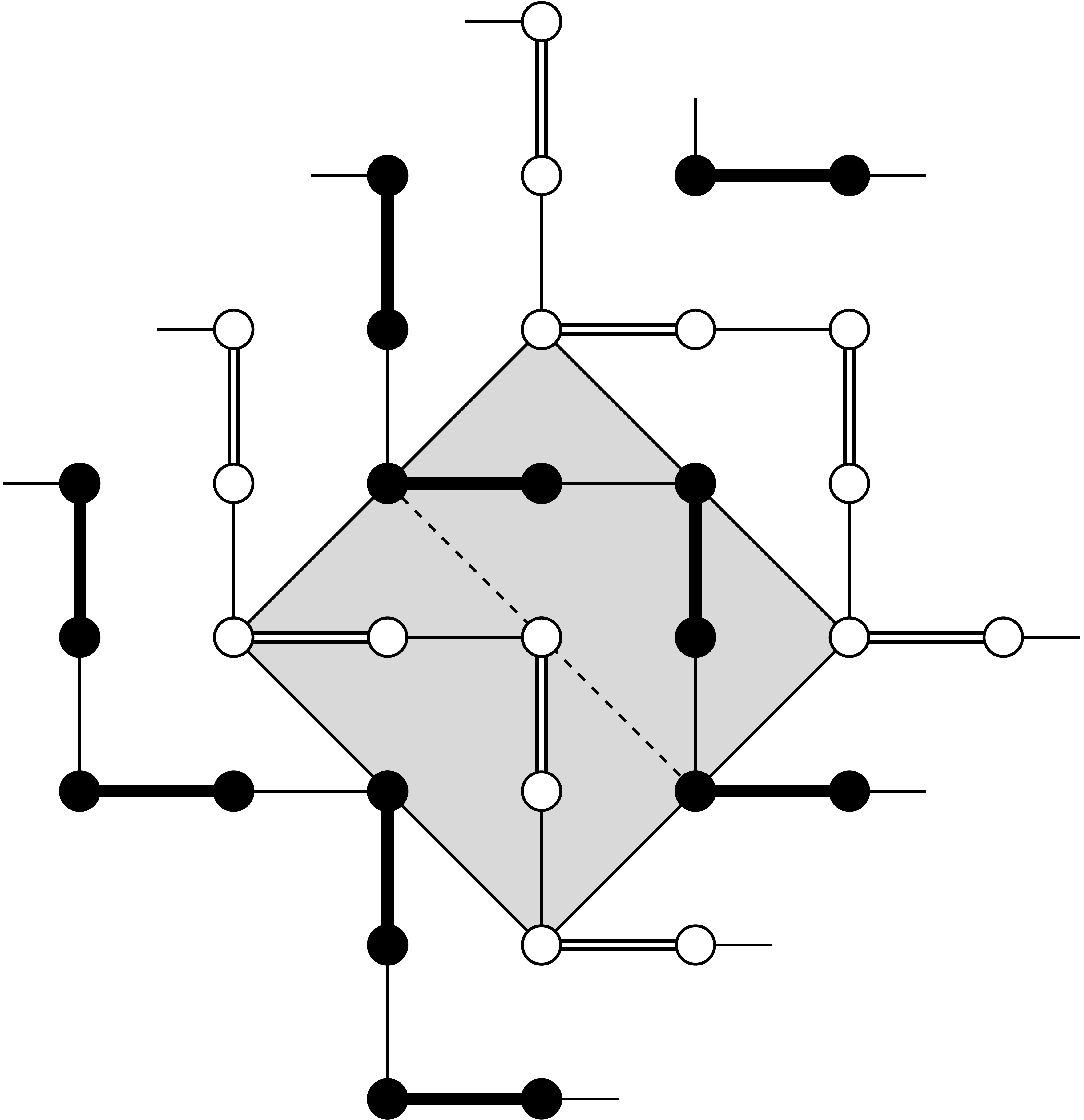}
\end{center}
\caption{\label{fig:afdimers}Schematic drawing of the Zener-polaron
  arrangement in the E-type (left) and the CE-type (right) magnetic
  order of {Pr$_{1/2}$Ca$_{1/2}$MnO$_3$}. Black and white spheres
  indicate the two spin orientations. Shown is the
  $ab$-plane. Different planes are antiferromagnetically coupled. The
  dimers are shown in the CE-structure as conceptual units, while the
  symmetry breaking in the real structure is small.}
\end{figure}

The prototypical order of Zener polarons at half filling is related to
the E-type magnetic structure shown in figure~\ref{fig:afdimers}. We
find this magnetic order higher in energy than the CE-type magnetic
structure. In the CE-type magnetic structure the Zener-polarons
polymerize so that the material exhibits one-dimensional zig-zag
chains of ferromagnetically coupled Mn ions.

\subsubsection{CE-type magnetic order}
\begin{figure}[!ht]
\begin{center}
\includegraphics[width=\linewidth]{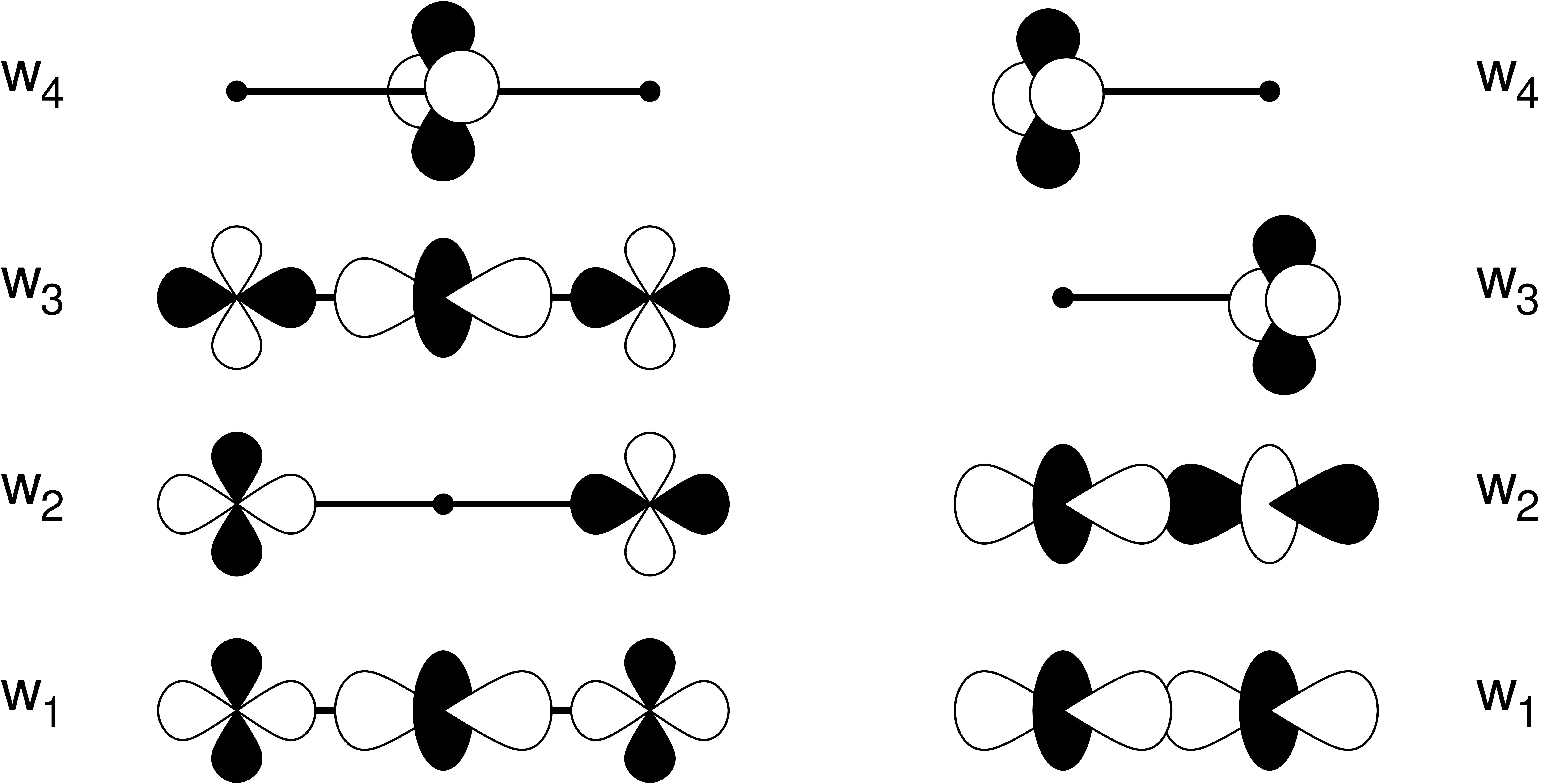}
\end{center}
\caption{\label{fig:ceorbs}Symmetry-adapted $e_g$ orbitals in the
  CE-type magnetic arrangement given in Eqs.~\ref{eq:defw1} and
  \ref{eq:w234} (left) and (right) the orbitals for a Zener
    polaron.  For both, the optical transition takes the electron
  from the corresponding orbital $|w_1\rangle$ to $|w_2\rangle$. }
\end{figure}

In order to understand the orbital ordering in the CE-type structure
it is helpful to consider a set of symmetry-adapted orbitals
constructed from majority-spin $e_g$ orbitals. Their schematic
sketches are shown in figure~\ref{fig:ceorbs}.

The first orbital $|w_1\rangle$ is a Wannier orbital for the occupied
$e_g$ band. It is localized on the three Mn ions forming one segment
of the ferromagnetically aligned zig-zag chain of CE-type
{Pr$_{1/2}$Ca$_{1/2}$MnO$_3$}.
\begin{eqnarray}
|w_1\rangle&=&|c,ax\rangle\cos(\alpha)
+\Bigl(|l\rangle
+|r\rangle\Bigr)\frac{\sin(\alpha)}{\sqrt{2}}
\label{eq:defw1}
\end{eqnarray}
For a three Mn-ion segment oriented in $x$-direction, the central
axial orbital $|ax\rangle$ is the $d_{3x^2-r^2}$ orbital on the
central atom.  The second, equatorial $e_g$ orbital on the central
site with $d_{y^2-z^2}$ character is denoted as $|eq\rangle$.

The $e_g$-orbitals at the corners of the segment are transformed into
a set of two orthonormal orbitals, $|l\rangle$ and $|r\rangle$, each
of which is oriented along one of the two segments of the zig-zag
chain. The resulting orbitals for the segment oriented along the
$x$-direction are
\begin{eqnarray}
|l\rangle= \Bigl(-|l,d_{3z^2-r^2}\rangle+|l,d_{x^2-y^2}\rangle
\Bigr)\frac{1}{\sqrt{2}}
\nonumber\\
|r\rangle= \Bigl(-|r,d_{3z^2-r^2}\rangle+|r,d_{x^2-y^2}\rangle\Bigr)
\frac{1}{\sqrt{2}}
\label{eq:lrofce}
\end{eqnarray}
The orbital rotated by 90$^\circ$ is obtained by flipping the sign of
the $d_{x^2-y^2}$-contribution in Eq.~\ref{eq:lrofce}.

The cubic harmonics are not aligned with the cartesian
coordinates, but are chosen consistent with the orientation of the
corresponding octahedron.

The remaining Hilbert space of $e_g$ orbitals of one segment of the
zig-zag chain is spanned by the following three orthonormal orbitals
\begin{eqnarray}
|w_2\rangle&=&
|l\rangle\frac{1}{\sqrt{2}}
-|r\rangle\frac{1}{\sqrt{2}}
\nonumber\\
|w_3\rangle&=&|ax\rangle\sin(\alpha)
-\Bigl(|l\rangle+|r\rangle\Bigr)\frac{\cos(\alpha)}{\sqrt{2}}
\nonumber\\
|w_4\rangle &=&|eq\rangle
\label{eq:w234}
\end{eqnarray}
shown in figure~\ref{fig:ceorbs}.

The value of the orbital-mixing angle $\alpha$ has been extracted from
the ab-initio DoS.  The best description of the valence band by the
orbital $|w_1\rangle$ alone is obtained with an orbital-mixing angle
$\alpha=30^\circ$.  This implies that the probability for the
electrons to reside on the central site is $3/4$ as opposed to $1/4$
on the corner sites.

The Jahn-Teller distortion $\sqrt{Q_2^2+Q_3^2}$ of the central Mn ion
with perfect orbital polarization is expected to scale about linearly
with charge. This is a direct consequence of Eq.~\ref{eq:optjtdist}.
For the calculated Jahn-Teller distortion of 0.259~\AA,
Eq.~\ref{eq:optjtdist} yields a net occupation of 0.63 electrons on the
central site.  A third, independent measure for the charge
disproportionation is the variation of the calculated magnetic
moments, which indicate that 0.7 electrons reside on the central atom
versus 0.3 electrons on the corner sites. Thus, our calculations
indicate charges $q=3.5\pm\Delta q$ with a charge-disproportionation
parameter $\Delta q$ in the range from 0.13~e to 0.25~e.

The measured charge disproportionation is even smaller than the
calculated ones. EELS line scan along the ordering axis in
{Pr$_{1/2}$Ca$_{1/2}$MnO$_3$} reveal that the variation of the Mn
valence state is below the measurement resolution,
i.e. $3.5\pm0.2$. However, a pronounced variation of the O $K$ pre-edge
intensity is found \cite{jooss07_pnas104_13597}. Valence-sum analysis
using the refined structure reveals a charge difference on two Mn
sites below 0.04. XANES studies of a full {Pr$_{1-x}$Ca$_{x}$MnO$_3$}
doping series reveal that the Mn spectra for $x=0.5$ can not be
described as a superposition of Mn$^{3+}$ and Mn$^{4+}$
spectra.\cite{mierwaldt14_catalysts4_129} For
Pr$_{0.6}$Ca$_{0.4}$MnO$_3$, no charge disproportionation has been
found albeit with fairly large error
bars,\cite{grenier04_prb69_134419} that are consistent with the
calculated values.

The origin of the charge order in the CE-type structure has been
attributed previously\cite{vandenbrink99_prl83_5118} to the onsite
Coulomb interaction between electrons in the $e_g$ shell. The origin
of this effect is that the Coulomb interaction, more precisely the
exchange term, favors orbital polarization. Because the $e_g$-orbitals
on the corner sites exhibit no orbital polarization, while the central
orbital does, the effective one-particle orbital on the central site
is lowered in energy. As a consequence it assumes a larger weight as
compared to the corner site.

A second effect, which acts in a similar fashion, has, however, not
been considered,\cite{vandenbrink99_prl83_5118} namely the
electron-phonon-interaction, respectively the Jahn-Teller
distortions. On the corner sites, both $e_g$ orbitals are equally
occupied, so that the orbital polarization is fully suppressed. In
contrast, the central site has complete orbital polarization, which
results in a pronounced Jahn-Teller distortion.

We estimate the resulting alternation of energy levels along the chain
from Eq.~\ref{eq:onsiteenergylevels} using our parameters. Using
occupations of $n_b=0.75$ for the central site and $n_a=n_b=0.125$ for
a corner site in the onsite model, we obtain an alternation of the
energy levels by 0.06~eV from the Coulomb interaction and 1.08~eV from
the Jahn-Teller effect. The estimate clearly attributes the charge
disproportionation to the  Jahn-Teller effect. The Coulomb
interaction plays a minor role, because in the manganites the two
Coulomb parameters cancel each other to a large degree.

Within the manifold of $e_g$ states, only the transition from
$|w_1\rangle$ to $|w_2\rangle$ has a non-vanishing dipole matrix
element. Transitions from the occupied orbital $|w_1\rangle$ to other
orbitals, on the same or other segments of the zig-zag chain, are
zero. The optically active transition can thus be identified with a
dipole oscillation between two corner Mn ions. The spectrum is shown
in figure~\ref{fig:optabsorb} alongside the experimental data.

The peak position at 1.5~eV is in agreement with the measured peak
position of 1.25~eV at 80~K of the optical conductivity due to the
near-infrared (NIR)
absorption.\cite{mildner15_prb92_35145} 

Interestingly, there is a fundamental difference in the nature of the
NIR absorption in the undoped manganite {PrMnO$_3$} and the one in the
half-doped material {Pr$_{1/2}$Ca$_{1/2}$MnO$_3$}. Whereas the former
can be attributed to an excitation from the lower to the upper
Jahn-Teller band of two neighboring Mn sites,\cite{kim06_prl96_247205}
the other is an excitation of a Zener polaron. To be precise, it is an
excitation within a ferromagnetic chain of polymerized Zener
polarons. The orbital order in $ab$-plane of {PrMnO$_3$} results in a
small delocalization of the bands and thus a small matrix element. The
dipole matrix element in the half-doped manganite on the other hand is
large in comparison. This is in accord with the measured doping
dependence of the optical conductivity.\cite{mildner15_prb92_35145}
The spectral weight increases from 125~eV/($\Omega$cm) to
300~eV/($\Omega$cm) from $x=0$ to $x=1/2$.\cite{mildner15_prb92_35145}

We obtain a fundamental band gap of 0.57~eV for the
{Pr$_{1/2}$Ca$_{1/2}$MnO$_3$} in the CE-type structure, and a direct
band gap of 0.72~eV.  All simple magnetic orders, namely G-type,
C-type, A-type and the ferromagnetic B-type order are energetically
unfavorable and do not exhibit a finite band gap in our calculations.

The calculated band gap of 0.57~eV for the half-doped material is
about 0.2~eV larger than experiment: Mildner et
al.\cite{mildner15_prb92_35145} obtained a band gap of 0.37~eV at 80~K
for doping $x=0.5$.  These values have been obtained from the
extrapolated onset of the optical conductivity of transitions between
Jahn-Teller split states (peak B in Mildner et
al.\cite{mildner15_prb92_35145}). The onset of optical conductivity at
reduced energy may be due to electronic transitions from phonon
excited states. The band gap due to Jahn-Teller splitting is larger
than the charge order gap at x=0.5, which has been estimated to be
0.15~eV at 80 K\cite{ebata07_prb76_174418} and 0.18~eV obtained at
10~K for $x=0.4$.\cite{okimoto98_prb57_9377}

\begin{figure}[htbp]
\begin{center}
\includegraphics[width=\linewidth]{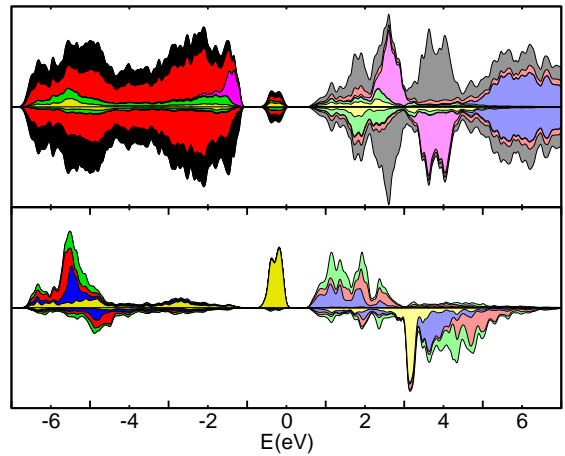}
\end{center}
\caption{\label{fig:pcmowannierdos}DoS of
  {Pr$_{1/2}$Ca$_{1/2}$MnO$_3$} in the stable CE-type
  antiferromagnetic order (top). The projection on symmetry-adapted
  orbitals (see figure~\ref{fig:ceorbs} and Eqs.~\ref{eq:defw1} and
  \ref{eq:w234}) is shown in the lower graph. The orbital
  $|w_1\rangle$ forming the occupied $e_g$ band is shown in
  yellow. The optically active excited orbital $|w_2\rangle$ is shown
  in blue. The fully antibonding orbital $|w_3\rangle$, shown in red,
  is optically inactive. The green area refers to the orbital
  $|w_4\rangle$ with $\delta$-symmetry on the central atom.}
\end{figure}

\section{Conclusion}
The electronic, magnetic and atomic structure of
{Pr$_{1-x}$Ca$_{x}$MnO$_3$} have been investigated with
first-principles calculations and experimental spectroscopy
studies. With the compositions $x=0, 1/2$, and $x=1$, we cover the
entire doping range. The comparison of experimental data with
calculated spectra provided additional insight far beyond the
capabilities of the individual techniques.

We use local hybrid functionals in our calculations that avoid the
well known deficiencies of density-functional calculations, which do
not account explicitly for the strong Coulomb interaction in the Mn-d
shell. By adjusting a small set of free parameters to Mn-XPS data,
good agreement with experimental spectra has been obtained for a large
spectral range. The parameter, we obtain for the admixture of the
Fock-term is substantially smaller than the value suggested on the
basis of perturbation theory.\cite{perdew96_jcp105_9982}

Specifically, the features in the low-energy spectral range of the
XANES and ELNES spectra, which is dominated by Mn-$e_g$ states
responsible for the correlation effects, could be assigned to specific
features of the calculated DoS. This opens new potential for analyzing
future high-resolution XANES and ELNES spectra with sub-100 meV energy
resolution.

Calculated band gaps lie within few tenths of an eV of the
experimental data. A local analysis of the calculated absorption
spectra attribute the NIR absorption band in {PrMnO$_3$} to a transition
from the lower to the upper Jahn-Teller band, where the electron is
excited into neighboring sites. For the half-doped material the
absorption is due to an internal excitation of a Zener polaron.

The large variation of the magnitude of the absorption coefficient for
upon reduction of {CaMnO$_3$} and the difference of the spectrum due to
charge-transfer transitions between {CaMnO$_3$} and {PrMnO$_3$} has
been traced back to the loss of spectral weight from the Mn-$e_g$
states, which become unavailable as final states for an optical
transition as they become occupied.

The measured optical absorption due to the NIR band in {PrMnO$_3$} is
substantially smaller than that in {Pr$_{1/2}$Ca$_{1/2}$MnO$_3$}. Our
calculations show that the NIR band in {PrMnO$_3$} depends strongly on
the magnetic order. For the A-type order the intensities for
{PrMnO$_3$} and {Pr$_{1/2}$Ca$_{1/2}$MnO$_3$} should be comparable.
The smaller optical absorption of {PrMnO$_3$} can be explained by
stronger antiferromagnetic correlations, such as in the G-type and
C-type structures. We anticipate that the observed state is a thermal
ensemble of several magnetic orders.

A detailed understanding of the absorption spectra and their
composition dependence is expected to guide the controlled preparation
of manganites.

Using the first-principles results we parameterized a model
total-energy functional. An important ingredient has been a consistent
description of the Coulomb interaction, because two independent
Coulomb parameters lead to a near-cancellation of the electron
interaction. This casts doubts on results based solely on one of the
Coulomb parameters. Coulomb interaction and electron-phonon coupling,
i.e. the Jahn-Teller effect, can lead to similar consequences in the
electronic structure. Our calculations indicate a dominance of the
Jahn-Teller effect over the Coulomb interaction within the Mn-$e_g$
orbital shell.

We find a considerable spatial variation of the hopping parameters,
which may be due to local variations of the Mn-O tilt angles in the
charge-ordered state.  Compared to models developed for {PrMnO$_3$},
we find rather small values for the antiferromagnetic coupling
$J_{AF}$, which we extract from {CaMnO$_3$}. The aggregate
antiferromagnetic coupling in {PrMnO$_3$} includes the super-exchange
between the $e_g$ electrons, which in the present model is treated
explicitly through the electronic wave functions.

The individual findings on the electronic structure and the
  excitations provide a sound basis for the understanding of energy
  conversion, optically driven phase transitions, and of the doping
  induced changes of properties, for example in the context of
  electrocatalysis.

\section*{Acknowledgements}
We thank Dr. Rosa Arrigo and Dr. Axel Knop-Gericke for fruitful
collaboration and support at ISISS beamline.  We further thank
Dr. Ulrich Vetter, Prof. Hans Hofs\"ass and Prof. Simone Techert for
providing technical equipment and support concerning the optical
absorption measurements. We thank Michael Ten Brink for carefully
reading the manuscript. Financial support from the Deutsche
Forschungsgemeinschaft (SFB 1073) through Projects B02, B03, C02 and
C03 is gratefully acknowledged.

\appendix
\section{Hamilton matrix elements from the generalized DoS}
\label{app:hfromdos}
Here we show how the hamilton matrix elements can be obtained from
the generalized DoS of a first-principles calculation.

\subsection{Hamiltonian of a non-orthonormal basis set}
Let $|\psi_n\rangle$ be the Kohn-Sham states of a density-functional
calculation and let $f_n$ be their occupation. The Kohn Sham states
are eigenstates of the Kohn-Sham Hamiltonian $\hat{H}$ with
energies $\epsilon_n$. 

We construct a basisset of local orbitals $|\chi_\alpha\rangle$ by
projecting the projector functions $|p_\alpha\rangle$ onto the
Kohn-Sham states in a given energy window $[a,b]$. We denote the set
of states with energies in the window by $M$.
\begin{eqnarray}
|\chi_\alpha\rangle=\sum_{n\in M} |\psi_n\rangle\langle\psi_n|p_\alpha\rangle
\end{eqnarray}

Hamilton and overlap matrix for this basisset can be expressed by the
generalized DoS, which has matrix elements
\begin{eqnarray}
D_{\alpha,\beta}(\epsilon)=\sum_n
\langle{p}_\alpha|\psi_n\rangle \delta(\epsilon-\epsilon_n)
\langle\psi_n|{p}_\beta\rangle
\;.
\end{eqnarray}

The matrix elements $H_{\alpha,\beta}$ of the Hamiltonian and those,
$O_{\alpha,\beta}$, of the overlap matrix elements are
\begin{eqnarray}
H_{\alpha,\beta}&=&\langle\chi_\alpha|\hat{H}|\chi_\beta\rangle
=\int_a^b d\epsilon\; D_{\alpha,\beta}(\epsilon)\epsilon
\label{eq:hamiltonfromdos}
\\
O_{\alpha,\beta}&=&\langle\chi_\alpha|\chi_\beta\rangle
=\int_a^b d\epsilon\; D_{\alpha,\beta}(\epsilon)
\label{eq:overlapfromdos}
\end{eqnarray}

\subsection{Hamiltonian of an orthonormal basis set}
In the model calculation, an orthonormal basisset is assumed.  Here, we
perform an approximate orthonormalization. 
\begin{eqnarray}
|\chi'_\alpha\rangle\approx
\sum_\beta |\chi_\beta\rangle ({\bm O}^{-\frac{1}{2}})_{\beta,\alpha}
\end{eqnarray}
  
The new Hamiltonian ${\bm h}$ in the orbital basis
  $\{|\chi'_\alpha\rangle\}$ is evaluated to first order in the
off-diagonal elements of Hamiltonian and overlap matrix.  This yields
\begin{eqnarray}
h_{\alpha,\beta}&\approx&
\frac{H_{\alpha,\beta}}{\sqrt{O_{\alpha,\alpha}O_{\beta,\beta}}}
-\frac{1}{2}\left(
\frac{H_{\alpha,\alpha}}{O_{\alpha,\alpha}}
+\frac{H_{\beta,\beta}}{O_{\beta,\beta}}\right)
\nonumber\\
&&\times\left(
\frac{O_{\alpha,\beta}}{\sqrt{O_{\alpha,\alpha}O_{\beta,\beta}}}
-\delta_{\alpha,\beta}\right)
\nonumber\\
&=&\epsilon_\alpha\delta_{\alpha,\beta}+t_{\alpha,\beta}
\label{eq:orthohamiltonfromdos}
\end{eqnarray}
where
\begin{eqnarray}
\epsilon_\alpha&=&\frac{H_{\alpha,\alpha}}{O_{\alpha,\alpha}}
\label{eq:orthohamiltonfromdosparmslevel}
\nonumber\\
\Delta_{\alpha,\beta}&=&
\frac{O_{\alpha,\beta}}{\sqrt{O_{\alpha,\alpha}O_{\beta,\beta}}}-\delta_{\alpha,\beta}
\label{eq:orthohamiltonfromdosparmshop}
\nonumber\\
t_{\alpha,\beta}&=&
\frac{H_{\alpha,\beta}}{\sqrt{O_{\alpha,\alpha}O_{\beta,\beta}}}
-\epsilon_{\alpha}\delta_{\alpha,\beta}
-\frac{\epsilon_\alpha+\epsilon_\beta}{2}\Delta_{\alpha,\beta}
\label{eq:orthohamiltonfromdosparms}
\end{eqnarray}
The diagonal elements of the matrix ${\bm t}$ of the hopping
parameters vanishes.

The Hamilton-matrix elements used to extract the hopping parameter
$t_{hop}$ are obtained from Eq.~\ref{eq:orthohamiltonfromdos} by
inserting Eqs.~\ref{eq:hamiltonfromdos} and
\ref{eq:overlapfromdos}.

\subsection{Splitting of onsite orbitals}
In order to extract the hopping parameters, we represent the orbitals
in a local coordinate system suitable for a particular oxygen bridge.
We consider the $d_{3z^2-r^2}$ orbitals with the local z-axis directed
towards the bridging oxygen atom.

The three onsite energies $\epsilon_\alpha$ obtained from the onsite
energies of the three octahedral axes contain the full information for
the onsite Hamiltonian of a given Mn-site. The quantity of interest
for the parameterization is the energy separation of the two onsite
levels, which enters the parameter determination as $\Delta^\uparrow$
and $\Delta^\downarrow$, respectively.

For each site we thus obtain the onsite hamiltonian expectation values
$\bar{h}_{i,i}$ for three non-orthonormal orbitals $d_{3x^2-r^2}$,
$d_{3y^2-r^2}$, $d_{3z^2-r^2}$ aligned with the three octahedral axes
from the corresponding level energies $\epsilon_\alpha$.  The bar
ontop of the symbol $\bar{h}_{ii}$ shall avoid confusion with the
matrix elements in the basis $\{|a\rangle,|b\rangle\}$.

The splitting $\Delta$ between the two
levels can be evaluated from the matrix elements $\bar{h}_{i,i}$ as
\begin{eqnarray}
\Delta&=&2\sqrt{2}
\sqrt{\frac{1}{3}\sum_{i=1}^3
\left[\bar{h}_{i,i}-\left(\frac{1}{3}\sum_{j=1}^3\bar{h}_{i,i}\right)\right]^2}
\end{eqnarray}
The matrix elements $\bar{h}_{ii}$ correspond to the level energy in
Eq.~\ref{eq:orthohamiltonfromdos} and
Eq.~\ref{eq:orthohamiltonfromdosparms}.

\section{Optical absorption expressed in local orbitals}
\label{app:optabs}
In order to relate the measured optical absorption spectra to specific
features in the DoS, it is desirable to express the
optical absorption in terms of local orbitals. This expression is
developed here.

Using second order perturbation theory in the limit of a nearly
monochromatic perturbation, we obtain the transition probability from
an initial state $|\psi_i\rangle$ with energy $\epsilon_i$ to a final
state $|\psi_f\rangle$ at energy $\epsilon_f$ as
\begin{eqnarray}
P_{f,i}&=&
\frac{2\pi}{\hbar}\Bigl|\langle\psi_f|W|\psi_i\rangle\Bigr|^2
\Bigl(1-f(\epsilon_f)\Bigr)f(\epsilon_i)
\delta(\epsilon_f-\epsilon_i-\hbar\omega)
\nonumber\\
\label{eq:transprobability}
\end{eqnarray}
where the perturbation of the hamiltonian is $H_1=W{\mathrm e}^{i\omega t}g(t)$
with a pulse shape $g(t)$. The pulse shape is normalized $\int
dt\;|g(t)|^2=1$. $f(\epsilon)=1/(1+\exp((\epsilon-\mu)/k_BT))$ is the
Fermi function.

The electromagnetic vector potential $\vec{A}$ and the scalar
potential $\Phi$ of a light pulse are described in the Coulomb gauge
as
\begin{eqnarray}
\vec{A}(\vec{r},t)
&=&
A_0\vec{e}_A{\mathrm e}^{i(\vec{q}\vec{r}-\omega t)}g(t)
\nonumber\\
\Phi(\vec{r},t)&=&0
\label{eq:lightpulse}
\end{eqnarray}
where $\omega=c|\vec{q}|$, $\vec{e}_A$ is a normalized polarization
vector and $\vec{q}$ is the dominant wave vector of the light pulse.
Some caution is needed due to the units we use for $A_0$ and $W$,
namely $\text{J}\sqrt{\text{s}}$ for $W$ and
$(\text{J/(Am)})\sqrt{\text{s}}$ for $A_0$.

The perturbation to first-order in the vector potential is
\begin{eqnarray}
W
&=&
-\frac{q}{2m}A_0
\Bigl((\vec{e}_A\vec{p}){\mathrm e}^{i\vec{q}\vec{r}}
+{\mathrm e}^{i\vec{q}\vec{r}}(\vec{e}_A\vec{p})\Bigr)
+O(|A_0|^2)\;.
\nonumber\\
\end{eqnarray}
We use  $\vec{e}_A=\frac{i}{\hbar}[\vec{p},(\vec{e}_A\vec{r})]_-$
to introduce the commutator with the unperturbed Hamiltonian
$H_0=\frac{\vec{p}^2}{2m}+v(r)$. After removing all
but the leading-order terms in the size $|\vec{q}|$ of the wave
vector, we obtain
\begin{eqnarray}
W
&=&
-\frac{q}{2m}A_0\frac{2mi}{\hbar}
\left[H_0,(\vec{e}_A\vec{r})\right]_- +O(|\vec{q}|,|A_0|^2)
\label{eq:pertyurbwithhamilkommut}
\end{eqnarray}

We express the wave functions in Eq.~\ref{eq:transprobability} in
terms of orthonormal local orbitals $|\chi_\alpha\rangle$ using
\begin{eqnarray}
|\psi_n\rangle=\sum_\alpha|\chi_\alpha\rangle c_{\alpha,n}
\label{eq:decomposewavefunc}
\end{eqnarray}
and we use the DoS matrix
\begin{eqnarray}
D_{\alpha,\beta}(\epsilon)
=\sum_n c_{\alpha,n}\delta(\epsilon-\epsilon_n)c^*_{\beta,n}\;.
\label{eq:dosmatrixforabsorption}
\end{eqnarray}

The matrix elements of the perturbation $W$,
Eq.~\ref{eq:pertyurbwithhamilkommut}, between local orbitals can be
evaluated using the hamilton matrix $h_{\beta,\alpha}$ from
Eq.~\ref{eq:orthohamiltonfromdos} as
\begin{eqnarray}
H_0|\chi_\alpha\rangle=\sum_\beta|\chi_\beta\rangle h_{\beta,\alpha}
\end{eqnarray}
as
\begin{eqnarray}
\langle\chi_\alpha|W|\chi_\beta\rangle
=
\frac{q}{i\hbar}A_0\sum_\gamma
&\Bigl(&h^*_{\gamma,\alpha}
\langle\chi_\gamma|(\vec{e}_A\vec{r})|\chi_\beta\rangle
\nonumber\\
&-&
\langle\chi_\alpha|(\vec{e}_A\vec{r})|\chi_\gamma\rangle 
h_{\gamma,\beta}
\Bigr)\;.
\label{eq:matrixelementw}
\end{eqnarray}

The absorption coefficient $a(\hbar\omega)$ is the ratio of absorbed
versus incident energy per unit thickness of the absorbing material,
\begin{eqnarray}
a(\hbar\omega)
&=&\frac{1}{\int dt\;\vec{S}(t)}\cdot\frac{1}{\Omega}\sum_{i,f\in\Omega}
P_{f,i}(\hbar\omega)\hbar\omega
\label{eq:abscoeffdef}
\end{eqnarray}
with the Poynting vector $\vec{S}(t)$. For the light pulse
of Eq.~\ref{eq:lightpulse}, the Poynting vector obeys
\begin{eqnarray}
\int dt\; \vec{S}(t)&=&\epsilon_0c|A_0|^2\omega^2\;.
\label{eq:integratedpoynting}
\end{eqnarray}
The sums over initial and final states are performed over all states in
the region $\Omega$ with volume $\Omega$.

Putting Eq.~\ref{eq:abscoeffdef} together with
Eq.~\ref{eq:integratedpoynting}, 
\ref{eq:transprobability},
\ref{eq:decomposewavefunc}, 
\ref{eq:dosmatrixforabsorption},
and
\ref{eq:matrixelementw}, we obtain the absorption
coefficient as
\begin{eqnarray}
a(\hbar\omega)
&=&
\frac{8\pi^2\alpha}{\hbar\omega}
\int d\epsilon\;
 (1-f(\epsilon+\hbar\omega))f(\epsilon)
\frac{1}{\Omega}\sum_{\alpha,\beta,\gamma,\delta\in\Omega}
\nonumber\\
&\times&
\sum_{u}
\Bigl(h_{\alpha,u}
\langle\chi_u|\vec{e}_A\vec{r}|\chi_\beta\rangle 
-
\langle\chi_\alpha|
\vec{e}_A\vec{r}|\chi_u\rangle 
h_{u,\beta}\Bigr)
\nonumber\\
&\times&
\sum_{v}
\Bigl(-h_{\gamma,v}
\langle\chi_v|
\vec{e}_A\vec{r}\Bigr)|\chi_\delta\rangle 
+
\langle\chi_\gamma|\vec{e}_A\vec{r}|\chi_v\rangle 
h_{v,\delta}\Bigr)
\nonumber\\
&\times&
D_{\beta,\gamma}(\epsilon+\hbar\omega)
D_{\delta,\alpha}(\epsilon)
\;.
\label{eq:abstb}
\end{eqnarray}
where we used $q=-e$ and the value of the dimension-less
fine-structure constant $\alpha=e^2/(4\pi\epsilon_0\hbar c)$.
Eq.~\ref{eq:abstb} is valid in the limits of low intensity, large
wave-length and slow variation of the pulse shape.

Now, we specialize the result to the Mn-$e_g$ orbitals. Assuming a
pure angular-momentum character and vanishing spatial overlap of
orbitals centered at different sites $\vec{R}_\alpha$, the matrix elements
simplify to
\begin{eqnarray}
\langle\chi_\alpha|(\vec{e}_A\vec{r})|\chi_\beta\rangle 
=(\vec{e}_A\vec{R}_\alpha)\delta_{\alpha,\beta}\;.
\label{eq:approxpureang}
\end{eqnarray}

In practice, it will be convenient to form a new basisset as
superposition of Mn-$e_g$ orbitals, so that
\begin{eqnarray}
D_{\alpha,\beta}(\epsilon)
=\sum_{a,b} q_{\alpha,a}\mathcal{D}_{a,b}(\epsilon)q^*_{\beta,b}
\;.
\end{eqnarray}
We obtain the final result
\begin{eqnarray}
a(\hbar\omega)
&=&
-
\frac{8\pi^2\alpha}{\hbar\omega}
\int d\epsilon\;
 (1-f(\epsilon+\hbar\omega))f(\epsilon)\frac{1}{V}\sum_{a,b,c,d\in\Omega}
\nonumber\\
&\times&
\Bigl[
\sum_{\alpha,\beta} q^*_{\alpha,a}
h_{\alpha,\beta}\vec{e}_A
\Bigl(\vec{R}_\beta-\vec{R}_{\alpha}\Bigr)q_{\beta,b}\Bigr]
\mathcal{D}_{b,c}(\epsilon+\hbar\omega)
\nonumber\\
&\times&\Bigl[
\sum_{\gamma,\delta}q^*_{\gamma,c}
h_{\gamma,\delta}
\vec{e}_A\Bigl(\vec{R}_\delta-\vec{R}_{\gamma}\Bigr)q_{\delta,d}\Bigr]
\mathcal{D}_{d,a}(\epsilon)
\end{eqnarray}

As the last step, we introduce the diagonal approximation by including
only contributions from the diagonal matrix elements of the DoS.  That
is, we only consider the diagonal elements of the DoS
$\mathcal{D}_{a,b}(\epsilon)$ in the new basisset.
\begin{eqnarray}
a(\hbar\omega)
&=&
\frac{8\pi^2\alpha}{\Omega}\sum_{a,b\in\Omega}\biggl|
\sum_{\alpha,\beta} q^*_{\alpha,a}
h_{\alpha,\beta}\vec{e}_A\Bigl(\vec{R}_\beta-\vec{R}_{\alpha}\Bigr)q_{\beta,b}
\biggr|^2
\nonumber\\
&\times&
\int \frac{d\epsilon}{\hbar\omega}\;
 \Bigl(1-f(\epsilon+\hbar\omega)\Bigr)f(\epsilon)
\mathcal{D}_{b,b}(\epsilon+\hbar\omega)\mathcal{D}_{a,a}(\epsilon)
\;.
\nonumber\\
\label{eq:absonsite}
\end{eqnarray}
This expression allows one to estimate the absorption probability from
a joint DoS. 

Note, however, that the diagonal approximation is accurate for a
basisset that already separates filled and empty states. If this
requirement is violated, a reasonable approximation would be to
perform again a sum over all orbital pairs, that is over all bonds,
but to include all contributions from that pair. The additional terms
depend on the off-diagonal terms of the DoS, i.e. the COOPs. In the
present paper this term is omitted, based on a careful choice of the
orbital basis.

The main approximations are
\begin{itemize}
\item the underlying independent-particle picture,
\item the assumption in Eq.~\ref{eq:approxpureang} that the orbitals
  have a pure angular momentum character of $d$-type and no spatial
  overlap, and
\item the diagonal approximation for the DoS matrix in
  Eq.~\ref{eq:absonsite}.
\end{itemize}
The main caveat of this local analysis of optical absorption is the
violation of momentum conservation, which is responsible for the
allowing transitions that preserve the Bloch vector. This, however,
turns into an advantage in strongly correlated materials, where
inhomogeneous fluctuations of the charge, orbital, spin, or lattice
degrees of freedom break translational symmetry, respectively provide
the momentum required for otherwise forbidden transitions. If the
long-range translational symmetry is broken, the coherence length of
the electrons is small, so that the local picture becomes valid.

The hamilton matrix elements have been evaluated using the model
parameters from section~\ref{sec:model}. The volume $\Omega$ has been
extracted from the measured lattice parameters of
Jirak\cite{jirak85_jmmm53_153} for {Pr$_{1/2}$Ca$_{1/2}$MnO$_3$}. The
Mn-Mn distance used to determine $\vec{R}_{\beta}-\vec{R}_\alpha$ is
the cube root of this volume.

\end{document}